\documentclass[preprintnumbers,twocolumn,groupedaddress,nofootinbib,aps]{revtex4}

\usepackage{amsmath,amssymb}
\usepackage{graphicx}
\usepackage{color}
\usepackage{hyperref}
\usepackage{url}
\usepackage{slashed}
\usepackage{subfigure}
\usepackage{slashed,bbm}
\usepackage{graphics,psfrag,epsfig}
\usepackage{dsfont}
\usepackage{enumerate}

\allowdisplaybreaks

\newcommand{\RNum}[1]{\uppercase\expandafter{\romannumeral #1\relax}}

\newcommand{\rund}[1]{ \left( #1 \right) }
\newcommand{\eck}[1]{ \left[ #1 \right] }

\newcommand{\pr}[1]{{{#1}^{\prime}}}

\newcommand{\bs}[1]{\boldsymbol{#1}}

\newcommand{\beq}{\begin{equation}}
\newcommand{\eeq}{\end{equation}}
\newcommand{\bea}{\begin{eqnarray}}
\newcommand{\eea}{\end{eqnarray}}
\newcommand{\be}{\begin{equation}}
\newcommand{\ee}{\end{equation}}

\begin{document}
\title{Interplay between magnetism, superconductivity, and orbital order in a 5-pocket model for iron-based superconductors -- a parquet renormalization group study }

\author{Laura Classen$^{1,2}$, Rui-Qi Xing$^{2}$, Maxim Khodas$^{3,4}$ and Andrey V Chubukov$^{2}$
}
\affiliation{$^{1}$ Institut f\"ur Theoretische Physik, Universit\"at Heidelberg, 69120 Heidelberg, Germany,\\
$^{2}$ School of Physics and Astronomy, University of Minnesota, Minneapolis, MN 55455, USA,\\
$^{3}$ Department of Physics and Astronomy, University of Iowa, Iowa City, IA 52242, USA,\\
$^{4}$  Racah Institute of Physics, The Hebrew University, Jerusalem 91904, Israel \\}

\begin{abstract}
We report the results of the
 parquet renormalization group (RG)
  analysis
   of the
  phase diagram  of the
  most general
  5-pocket model for Fe-based superconductors.
 We
 use as an input the orbital  structure of
  excitations near
  the
   five pockets made out of $d_{xz}$, $d_{yz}$, and $d_{xy}$ orbitals and
  argue that there are 40 different interactions between low-energy fermions in the orbital basis.
     All interactions flow
      under RG, as one progressively integrates out fermions with higher energies.
       We find that the low-energy behavior is amazingly simple,
       despite the large number of interactions. Namely,
        at low-energies the full 5-pocket model effectively reduces either to a 3-pocket model made of one $d_{xy}$ hole pocket and two electron pockets,
         or a 4-pocket model made of two $d_{xz}/d_{yz}$ hole pockets and two electron pockets.
          The leading instability in the effective 4-pocket model is a spontaneous orbital (nematic) order, followed by $s^{+-}$ superconductivity.
          In the effective 3-pocket model orbital  fluctuations are weaker, and the system develops either  $s^{+-}$ superconductivity or stripe SDW.
          In the latter  case, nematicity is induced by composite spin fluctuations.
\end{abstract}

\maketitle

{\it \bf Introduction.}~~~The  interplay between superconductivity,
 magnetism, and nematicity  is the key
 %element in   to the
 physics of Fe-based superconductors (FeSCs)
  ~\cite{review,rev_1,orb_order,ch_rev,  we_last,zlatko}.
In some FeSCs, e.g., 1111 and 122  systems,
 undoped materials display a stripe magnetic order below a certain $T_N$ and a nematic order at slightly higher temperatures, while
  superconductivity emerges upon doping, when magnetic order gets weaker.  In other systems, like 111 LiFeAs and 11 FeSe,
  superconductivity emerges without long-ranged magnetism already in undoped systems. Besides, FeSe displays an orbital order above the superconducting (SC) $T_c$ \cite{FeSE}.
 The issue for the theory is to understand whether these seemingly different behaviors can be understood within the same framework.

In this communication we report the results of our analysis,
which connects
different classes of FeSCs.
 We
 %use an itinerant approach, by which we mean that {\it low-energy}
% fermionic states are itinerant, and
study the competition between superconductivity,
 magnetism, and nematicity in the  most generic five-pocket (5p)  model for FeSCs with full orbital content of low-energy excitations.
  To do this, we  use the machinery of analytical parquet renormalization group (pRG)~\cite{parquet_RG}.
    This approach, along with complementary
    %AC
     numerical
     functional RG~\cite{fRG,fRG_thomalle,fRG_thomale2,fRG_lee},
  has been argued~\cite{fRG,rice,rahul,fRG_thomalle,fRG_thomale2,fRG_lee,cee,pod,we_last,kontani}
   to be the most unbiased way to
 analyze competing orders
   in an itinerant electron system.

\begin{figure}[t!]
\centering
 \includegraphics[width=1\columnwidth]{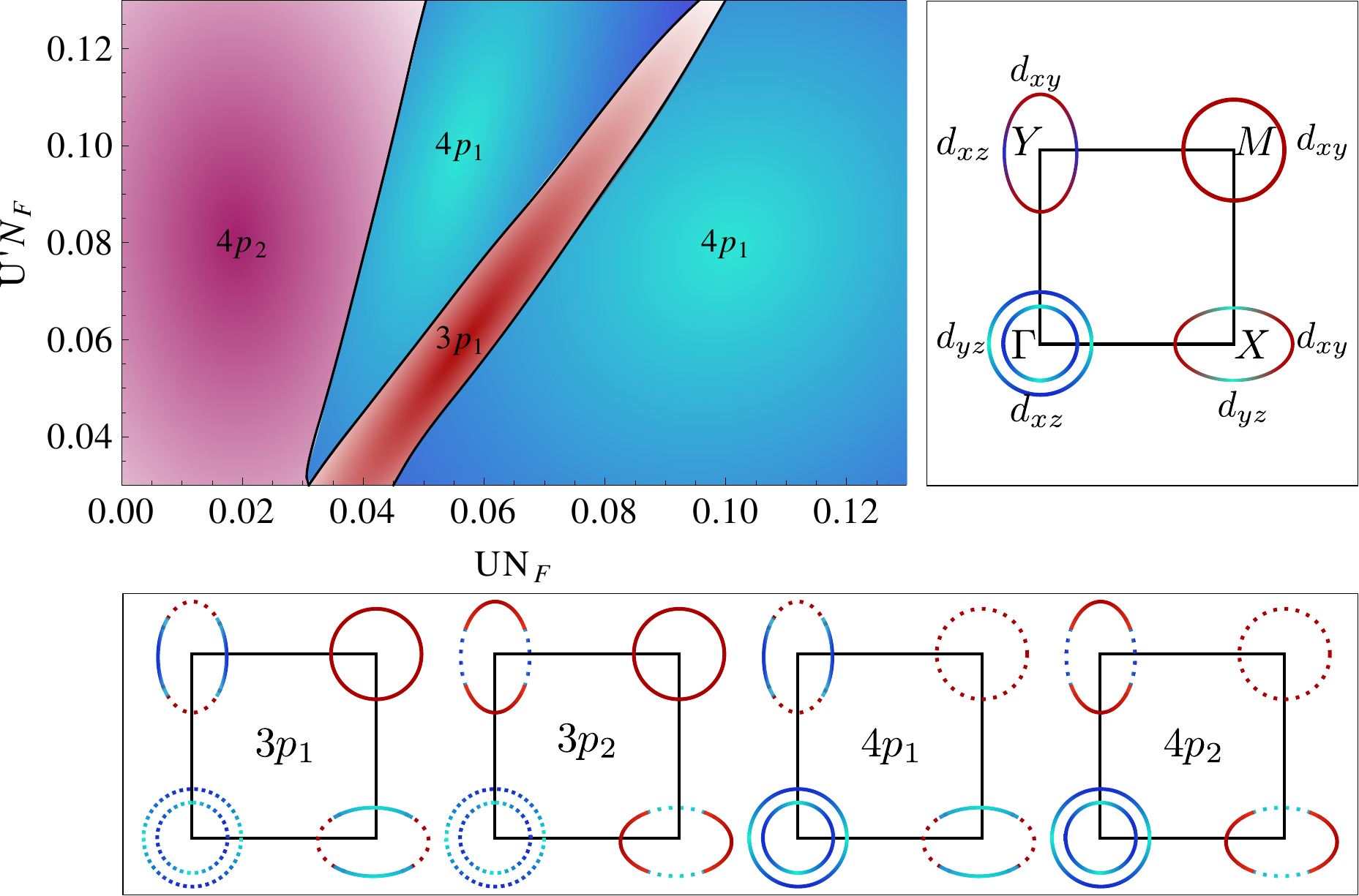}
 \caption{
  Upper panel: Right -- main orbital content of excitations near Fermi surfaces (presented by different colors).  Left -- regions of different system behavior of the full 5-pocket model, indicated by the type of the effective model.
  In the ranges marked 3p$_{1,2}$,
    the dominant interactions at low energies are within the subset of the two electron pockets and the $M=(\pi,\pi)$-hole pocket. In the ranges marked 4p$_{1,2}$,
   the dominant interactions are between  fermions near the $\Gamma$-centered hole pockets and electron pockets.  The index $1,2$ distinguishes if interactions
    involving $d_{xz}/d_{yz}$ or $d_{xy}$ orbital components on the electron pockets are dominant.
    For illustrative purposes, 
    the bare model is set to have local Hubbard and Hund interactions -- intraorbital $U$, interorbital $U'$,  $J$ and $\pr{J}$. We set  $J=0.025/N_F$, $\pr{J}=0.03 /N_F$, where $N_F$ is the density of states on the FSs (assumed to be equal on all FSs for simplicity), and varied $U$ and $U'$ as two independent parameters.
     Lower panel:  Graphic representations of  3p$_{1,2}$ and  4p$_{1,2}$ models.  Fermionic states, for which interactions become the largest in the process of pRG flow,
      are shown by solid lines. }
    \label{fig:full}
\end{figure}

 %We consider the most generic five-pocket (5p) model for FeSCs,with
  The 5p model consists of three hole pockets, of which two are centered at $\Gamma=(0,0)$ in the 1Fe Brillouin zone and one is centered at $M=(\pi,\pi)$, and two electron pockets centered at $Y=(0,\pi)$ and $X=(\pi,0)$
(see the right panel in Fig.~\ref{fig:full}).
    The two $\Gamma$-centered hole pockets are made out of $d_{xz}$ and $d_{yz}$ orbitals,
  the hole pocket at $M$ is made out of
     $d_{xy}$ orbitals. The electron pockets are made out of $d_{xz} (d_{yz})$ and $d_{xy}$ orbitals
     ~\cite{scalapino, brouet}.

 For such an electronic configuration, there are
 40 different 4-fermion interaction terms, allowed by $C_4$ symmetry~\cite{sm_1}  (without the hole pocket at $M$, this number is 30~\cite{Cvetkovic2013}).
 If one departs from the model with only local interactions, the
  bare values of all 40  interactions are linear combinations of inter- and intra-orbital Hubbard and Hund terms $U$, $U'$, $J$ and $J'$.
 However, the 40 interactions flow to different values under pRG, which implies that the system self-generates non-local interactions.
The flow of the interactions is obtained by solving differential equations that encode series of coupled vertex renormalizations.
%As a second step within pRG, the
The running interactions are then
used as input to determine susceptibilities in different channels.
% A divergence of a susceptibility upon lowering of
%energy/temperature signals the development of an ordered phase.
This way one can monitor
%the
 a simultaneous build-up of different correlations taking into account their mututal feedback, which turns out to be crucial in our study.
% for FeSCs.

   The main
   %advantage
    result of pRG
    analysis
     is the emergent
   universality. It means that
   % although there are
   40
   microscopic
   interactions
   %, they
   flow towards a {\it limited number} of fixed trajectories (FT), where the ratios of different interactions become universal numbers.
      Each fixed trajectory has a basin of attraction in the space of bare interaction parameters.
      %, i.e., microscopic models  flow towards one or the other FT depending on $U, U', J, J'$.
         This allows us to explain the rich behaviors of the different FeSCs within a unifying description.
    In practical terms
    a  simultaneous build-up of different correlations
   % , the pRG approach  relies on two-dimensionality, on parabolic dispersion of excitations around $\Gamma$, $M$, $X$ and $Y$ points, and on the presence
 % of both hole-like and electron-like excitations. It is applicable when bare interactions are, at most, comparable to the bandwidth, $W$,  and  pRG equations hold
  holds in the window of energies between a fraction of $W$ and a scale comparable to the  Fermi energy, $E_F$. At smaller energies, interactions in different channels evolve independent on each other.
    The range between $W$ and $E_F$ should be wide enough, otherwise the pRG flow ends  before the system reaches one of the FTs
     %(the subsequent flow of couplings  is  RPA-like
     ~\cite{maiti}.
     %).

{\it \bf Summary of our results.}~~~
    % Despite that the number of interactions is large, we found only
     We found
      four stable FTs.
     For the  first  two stable FTs, the interactions within the  subset of the two $\Gamma$-centered hole pockets and the two electron pockets become dominant, i.e., the 5p model effectively reduces to the four-pocket model (4p).
          For the other two stable FTs, the 5p model reduces to an effective 3-pocket model (3p) consisting of  two electron pockets and the $M$-hole pocket.
            On each of two stable 4p FTs or 3p FTs the system behavior is  described by an even simpler effective model because interactions  involving fermions from
             either $d_{xz}/d_{yz}$ or $d_{xy}$ orbitals become dominant.  We label these models as  4p$_1$,  3p$_1$, and 4p$_2$, 3p$_2$, respectively.
    We illustrate the four cases and present the phase diagram in Fig. \ref{fig:full}.  We then  computed susceptibilities in different channels
    \cite{suscept_RG}.  We found that the interplay between spin-density-wave (SDW) magnetism and superconductivity is the same in all four effective models.
    Namely, the SDW susceptibility is the largest at intermediate energies  and pushes SC and orbital susceptibilities up. However, in the process of the pRG flow the SC susceptibility overtakes the SDW one,
    and the feedback from SC fluctuations halts the increase of the SDW susceptibility
    (see Fig. \ref{fig:ias}(b)).
     As a consequence, already the undoped system develops superconductivity rather than SDW magnetism, if indeed the pRG flow runs over a wide enough range of energies.
    This result could not be obtained within RPA and is entirely due to the feedback from increasing SC fluctuations on the SDW channel.
     In all cases superconductivity is of $s^{+-}$ type, with sign change between the gaps on hole and electron pockets.
       In
       %both
        4p models  the susceptibility towards $C_4$-breaking orbital
       order  also grows, and its exponent is larger than that for superconductivity~\cite{we_last}, i.e., the system
       first develops a spontaneous orbital order. In 3p models orbital fluctuations are much weaker,
        and orbital order does not have
         enough
          "space" to develop.

 We found that SDW magnetism does
      develop before superconductivity and/or orbital order if the FT is not reached within the range of pRG flow.
       The type of  SDW order is different for the 3p and the 4p models.  In 3p models SDW order is a  $C_4$-breaking stripe order~\cite{eremin,sdw_stripe},
       while in 4p models it is  $C_4$ preserving double-Q order~\cite{fkc,raf_last}
       %, which is
       (a symmetric combination of $(\pi,0)$ and $(0,\pi)$ magnetic orders).  This last result, in combination with pRG, implies a clear separation
       between the magnetic and orbital scenario
        for nematicity in FeSCs.  Namely, in 4p models, the SDW scenario for Ising-nematic  order
         does not work because double-Q SDW preseves the symmetry between $X$ and $Y$ directions,
         and, simultaneously, orbital fluctuations are strong. In 3p models,
          orbital fluctuations are weak,   and, simultaneously, SDW {\it stripe} fluctuations favor vestigial Ising-nematic spin order~\cite{rafael}.
       \begin{figure}[t!]
\centering
 \includegraphics[width=.75\columnwidth]{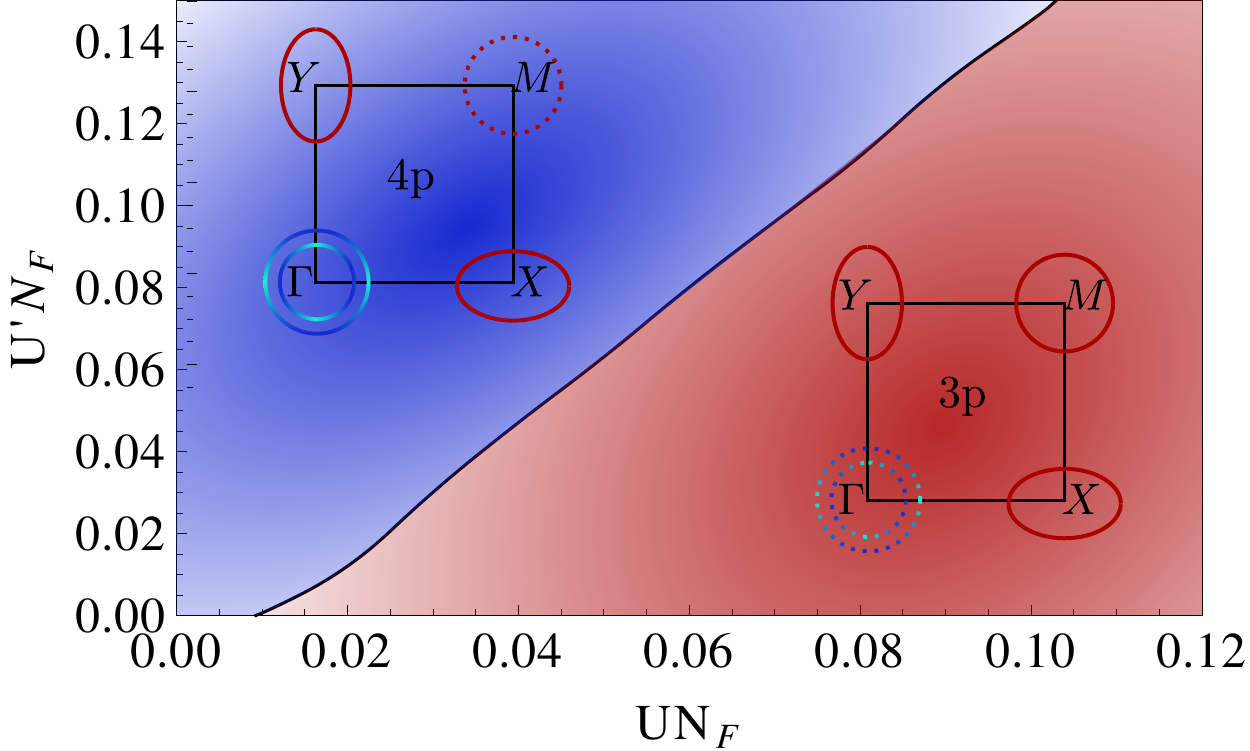}
 \caption{
 Two different regions of system behavior indicated by fixed trajectories of the pRG flow for the toy model with electron pockets made entirely of $d_{xy}$, for different values of $U, U'$
  (treated as two independent parameters) and $J=\pr{J}=0.03 /N_F$.  In the region labeled as 3p the  interactions  within the subset of the two electron pockets and the $M=(\pi,\pi)$-hole pocket become dominant at low energies. In the region labeled as 4p
  interactions involving fermions from the two $\Gamma$-centered hole pockets and the two electron pockets become  dominant.}
   \label{fig:simpl}
\end{figure}

In the remainder of this Letter we present the details of our study.
  The full analysis of the set of 40 pRG equations is quite involved, so to demonstrate the separation into 4p or 3p behavior at low energies, we first analyze a toy  model, in which we approximate the orbital composition of the two electron pockets as pure $d_{xy}$. We then extend the analysis to the full 5-pocket model.

 {\it \bf Toy model with $d_{xy}$ electron pockets.}~~~
 As we said, the
 kinetic term describes fermionic excitations around the five Fermi surfaces, i.e. $H=H^\Gamma+H^X+H^Y+H^M$.
 The  symmetry-allowed interaction terms
  % are determined like it is done in a g-ology \cite{giamarchi}
  %  and
    contain 14 interactions
     $U_i$
    within the subset of the two electron and the two $\Gamma$-centered hole pockets and 7 interactions $U_{in}$ involving fermions near the $M$-hole pocket, so the total number
    of the interactions is 21.
     We present the Hamiltonian and the full set of pRG equations  for a generic dispersion near hole and electron FSs in the Supplementary Material (SM).
     %,
    % and here for simplicity present the results assuming that the density of states $N_F$ is the same on all  pockets.
      The
      pRG
      analysis shows that six interactions flow to zero and five increase with smaller exponents than the other ten.
     The pRG flow of the remaining ten interactions determines the FTs.  We show these ten interactions
      in the inset of  Fig. \ref{fig:ias}(a).
      The pRG equations for these interactions are ($u_i = U_i N_F$)
\bea
     \label{eq:rgxy}
&&\dot u_1 =u_1^2+u_3^2,~\dot u_{1n}=u_{1n}^2+u_{3n}^2 \\
&&\dot u_2 =2u_2(u_1-u_2),~\dot u_{2n}=2u_{2n}(u_{1n}-u_{2n}) \nonumber \\
&& \dot u_3=2u_3(2u_1-u_2-u_5)-2  u_3u_{4}-u_{3n}u_{5n} \nonumber\\
&&\dot u_{3n}=2u_{3n}(2u_{1n}-u_{2n}-u_5)-u_{3n}u_{4n}-2 u_3 u_{5n} \nonumber \\
&&\dot u_4 =-2 u_4^2-2u_3^2-2u^2_{5n},\dot u_{4n}=-u_{4n}^2-2u_{3n}^2-2 u^2_{5n} \nonumber\\
&&\dot u_5=-2u_5^2-2 u_3^2-u_{3n}^2, \nonumber\\
&&\dot u_{5n}=-2u_4u_{5n}-u_{4n}u_{5n}-2u_3u_{3n}\nonumber
\eea
The derivatives are  with respect to $L = \log W/E$, where $E$ is the running scale.

We searched  for FTs of Eq. (\ref{eq:rgxy}) by selecting one divergent interaction (specifically $u_1$ or $u_{1n}$), writing other interactions as
 $u_i = \gamma_i u_1$,  $u_{in} = \gamma_{in} u_{1}$ (or $u_i = \gamma_i u_{1n}$,  $u_{in} = \gamma_{in} u_{1n}$), and solving the set of equations for
  $L-$independent
  $\gamma_i, \gamma_{in}$.
   We found two stable FTs:
    one with
\bea
&&u_1=\frac{1}{1+\gamma^2_{3}}\frac{1}{L_0-L},
\label{a}
\eea
and
 $\gamma_{in} = \gamma_2 = 0$, $\gamma_3=\pm \sqrt{15}, \gamma_4=\gamma_5=3$,
 and the
 other with
\bea
&&u_{1n}=\frac{1}{1+\gamma_{3n}^2} \frac{1}{L_0-L}
\label{b}
\eea
and $\gamma_{1}=\gamma_{2}=\gamma_{3}=\gamma_{4}=\gamma_{2n}=\gamma_{5n}=0$, $\gamma_{3n}=\pm (3+2\sqrt{6}), \gamma_{4n}=2\gamma_5= -\sqrt{6}$.
In Eqs. (\ref{a}), (\ref{b}) $L_0$ is the scale at which interactions diverge and the system develops a long-range order, as we show below.
For the first stable FT all $\gamma_{in}$ involving the $M$ pocket vanish, so the 5-pocket model effectively reduces to the 4p model.
 For the second stable FT the situation is the opposite -- interactions involving the two $\Gamma$-centered hole pockets vanish compared to other interactions,
 i.e., the 5p model effectively reduces to the 3p model.
     We checked the stability of the 4p FT and the 3p FT by expanding around them and
    verified that all eigenvalues
    are negative.
  Whether the system flows to one FT or the other is  determined by the bare values of the interactions (see Fig.\ref{fig:simpl}).

We next
use the running interactions as inputs and compute the susceptibilities in different channels, $\chi_j$.
   We describe the computational procedure in the SM and here list the results.
    The
    potentially divergent
    parts of the
    susceptibilities in  SC and SDW channels are $\chi_i \propto (L_0-L)^{2\beta_i-1}$ (i = SDW, SC).
 Along 4p FT and 3p FT, the
 %largest
  exponents are $\beta_{SDW}^{(4p)}= 0.30, \beta_{SC}^{(4p)}= 0.86, ~\beta_{SDW}^{(3p)} =  0.43, \beta_{SC}^{(3p)} = 0.72$.
 We  see that in both cases
   $\beta_{SC} >1/2$ while $\beta_{SDW} <1/2$, i.e.
    $\chi_{SC}$ diverges at $L=L_0$, while $\chi_{SDW}$ remains finite, despite that it was the largest at the beginning of the pRG flow.
     This implies that
    the system develops SC order but not SDW order.
    We show the flow of the susceptibilities
     in Fig.~\ref{fig:ias}(b).
For both 4p and 3p models, we found that the largest $\beta_{SC} >0$ corresponds to the $s^{+-}$  gap structure, with opposite sign of the gap on hole and electron pockets
~\cite{comm_kot}

\begin{figure}[t!]
\centering
 \includegraphics[width=.7\columnwidth]{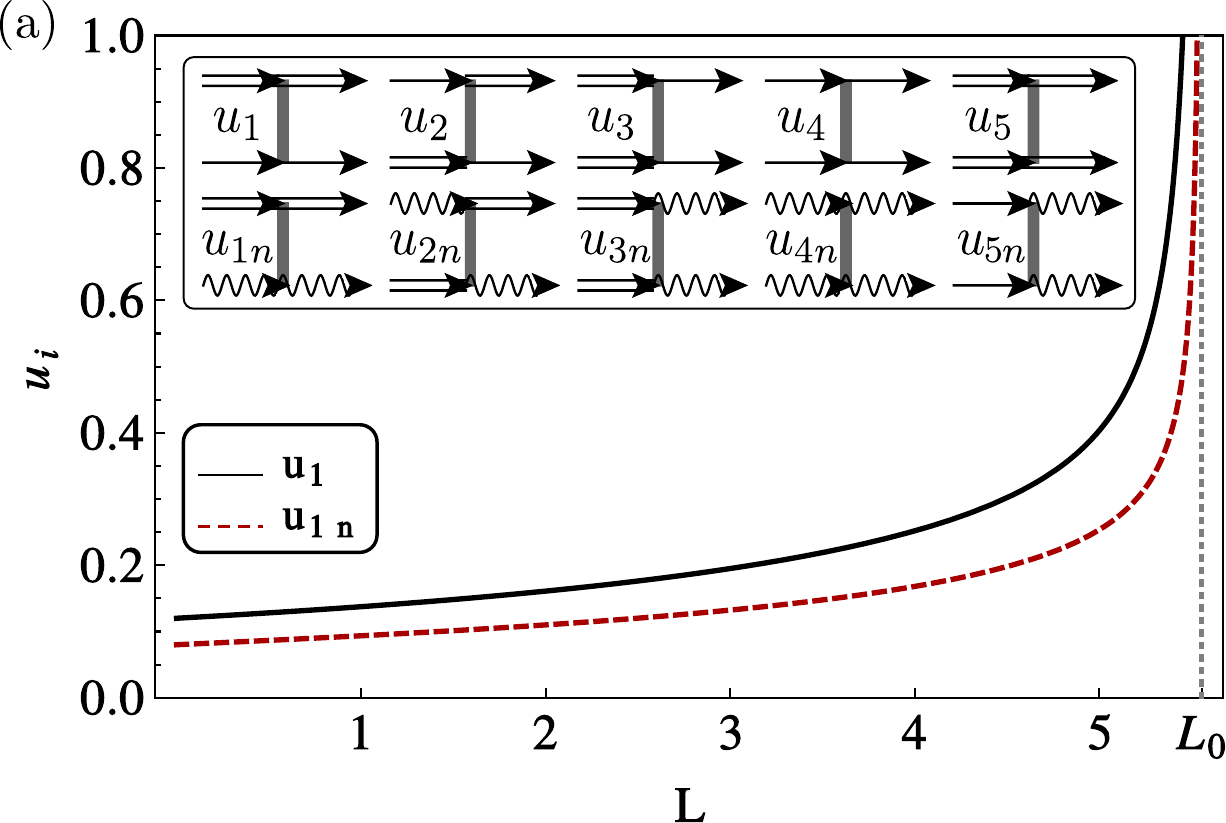}
 \\
 \includegraphics[width=.7\columnwidth]{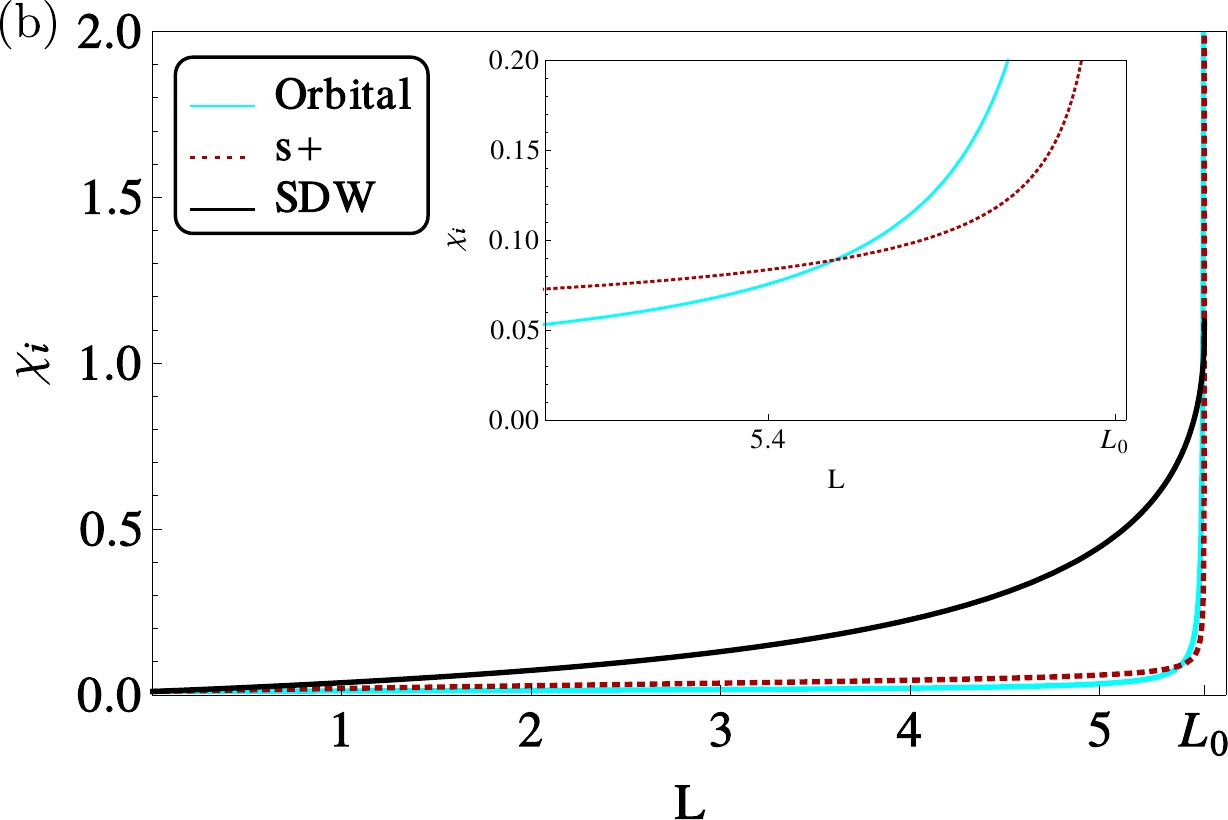}
 \caption{(a) Representative RG flow towards the 4p FT in the toy model for the interactions $u_1$ and $u_{1n}$. The inset shows the 10 relevant interactions of the toy model, where double lines represent electron pockets, wavy lines the $M$-centered hole pocket and solid single lines the $\Gamma$-centered hole pockets. Bare values are $U=0.08/N_F$, $U'=0.12/N_F$, $J=\pr{J}=0.03/N_F$. The RG parameter $L$ is $\log{W/E}$, where $W$ is the bandwidth and $E$ is running energy/temperature.   The system undergoes an instability into an ordered state (SDW, SC, or orbital order) at $L = L_0$.
  (b) Corresponding flow of the SDW, SC $s^{+-}$ and
   orbital  susceptibilities. Near $L = L_0$  the  SC  and the orbital susceptibilities  keep increasing, while the SDW susceptibility remains finite.
   %LC
   The inset shows orbital and SC susceptibilities at the end of the flow.}
   \label{fig:ias}
\end{figure}

 We also analyzed the susceptibility $\chi_P$ in the
    $d$-wave Pomeranchuk channel. An instability in this channel
     leads to spontaneous orbital order~\cite{orb_order,we_last},
      i.e., non-equal densities of fermions on $d_{xz}$ and $d_{yz}$ orbitals.  For the 4p model we
     found that
$\beta_{P}^{(4p)} =1$ is larger than
$\beta_{SC}^{(4p)}$, i.e., orbital order can precede the SC transition~\cite{we_last}.
 We found no
  $d_{xz}/d_{yz}$ orbital order for the 3p model because the electron and the $M$ pockets have $d_{xy}$ character~\cite{oo}.

{\it \bf Full 5-pocket model.}~~~ The analysis of the full 5-pocket model with $d_{xz}/d_{xy}$ and $d_{yz}/d_{xy}$ orbital content of the electron pockets is more involved as one has to analyze the set of 40 coupled differential equations for the interactions (see SM).
    We searched for FTs with the same procedure as in the toy model.
    Amazingly enough, we found much the same behavior.
    Namely, the 5p model effectively becomes either a 4p  or a 3p model.
     The new feature, not present in the toy model, is that in each case there are now two stable FTs,  on which the system behavior is described by even more restricted
      $3p_{1,2}$ and $4p_{1,2}$ models.  For $3p_1$ and $4p_1$ models
     interactions involving fermions from $d_{xz}$ ($d_{yz}$) orbitals on the electron pockets become dominant, for  $3p_2$ and $4p_2$ models
     interactions of $d_{xy}$  orbitals on the electron pockets become dominant.
        We verified that these four FTs are stable
        %in the  full  5p model.
         with respect to small deviations.
           We show the phase diagram in Fig.~\ref{fig:full}.

   The interplay between SDW and  SC  is the same in all four effective models and is similar to that in the toy model.
 Namely,
 the SDW susceptibility is the largest at the beginning,
   but in the process of the flow SC susceptibility diverges faster, and the feedback from SC fluctuations halts the growth of SDW susceptibility.  As a result, even at zero doping the system develops $s^{+-}$ SC order but no SDW order.
    Orbital fluctuations are, however, different in 4p and 3p models, again in similarity to the toy model.
      If the pRG flow is towards  $4p_1$ or $4p_2$ models, orbital
      %(d-wave Pomeranchuk)
       fluctuations also get strong and $\chi_P$ diverges with the largest exponent, i.e., the system
       % may develop
        develops a spontaneous orbital order prior to SC \cite{we_future}.
      If the flow is towards 3p model, orbital fluctuations are
       much weaker
       and do not develop for not too large $W/E_F$.
If
%the  largest
$E_F$ is larger than $E_0 \sim We^{-L_0}$,
the pRG flow ends
  before $\chi_{SC}$ and/or $\chi_P$  wins over $\chi_{SDW}$.
  In this situation, the system
  %behavior is determined by doping, which affects the behavior of $\chi_{SDW} (E)$ below $E_F$. In a generic case
% the system
 develops SDW order at smaller doping and SC order at larger dopings~\cite{maiti}.
For the 4p model an SDW order is a
  double-Q order,
  maintaining the symmetry between $X$ and $Y$ directions\cite{fkc,raf_last}, while  for the  3p model SDW order is a
    stripe,
    % order,
    breaking this symmetry.
    ~\cite{eremin,sdw_stripe}.
  Combining this with pRG results, we find  that, if the pRG flow is towards one of the two 4p models, the
  %only option for
  nematicity emerges as
  %is
   a spontaneous orbital order. If the flow  is towards one of the 3p models,
   %orbital fluctuations are weak, but now the system develops
    the nematicity emerges due to stripe fluctuations  as a composite  Ising-nematic spin order.
    % vestigial to the stripe SDW order.

{\it \bf Applications to FeSCs.}~~~Our results have several implications for FeSCs.  First, the pRG analysis shows that
   SC order may develop instead of long-ranged magnetism already in undoped materials, not only when SDW order is destroyed by doping. This is consistent with the behavior in
  LiFeAs and FeSe \cite{comm_1}.
     In systems with smaller regions of the pRG flow (larger bare interactions or larger $E_F$) SDW order develops first, and SC develops only upon doping.
   Second, pRG analysis shows that in 4p models  orbital order develops first,  SC  develops at a lower $T$, and SDW order does not develop down to $T=0$. This is consistent with the observed behavior in FeSe at ambient pressure \cite{FeSE}.
      The third result is the separation between orbital and magnetic scenarios for nematicity
      in 4p and 3p models.
       %In 4p models, orbital fluctuations are strong and, simultaneously, SDW fluctuations favor $C_4$-symmetric checkerboard SDW.  In this situation, nematicity is the result of a spontaneous orbital order.
      %In 3p models,  orbital fluctuations are weak, and, simultaneously, SDW fluctuations favor $C_4$-breaking stripe SDW.
      %  In this situation,  nematicity is caused by Ising-nematic composite spin order, which precedes stripe SDW~\cite{rafael}.
        Whether the system flows towards 3p or 4p effective model at low energies depends on the microscopic Hubbard and Hund  parameters  (see Figs.~\ref{fig:full},~\ref{fig:simpl})
       as well as the parameters of fermionic dispersions (see SM).

{\it \bf Conclusions.}~~~ In this Letter we analyzed the competition between SDW, SC, and orbital order
 in the
full 5-pocket model for FeSCs.
We used  pRG techniques and included into consideration the
 orbital composition of hole and electron pockets
  in terms of $d_{xz}, d_{yz}$, and $d_{xy}$  orbitals.  The total number of
   symmetry-allowed interactions between low-energy fermions  is 40,
     yet we found
    the
     system behavior
     is  amazingly simple -- depending on initial values of the
  interactions  and  quasiparticle masses
   the system flows to one of four stable FTs.  For two of these FTs,
   the system behavior at low energies
    is the same as
     if the the M-pocket was absent (4p model), for the other two the
     system behavior
     is the same as if the two $\Gamma$-centered hole pockets were absent (3p model).
       In
        all
        cases
         $s^{+-}$ SC wins over SDW
         if $E_F$ is small enough, and SDW wins if $E_F$ is larger.
        In the parameter range where the pRG flow is towards the effective 4p model,
         the system develops
         %  first a
         spontaneous orbital order, which
          then
          is the origin of nematicity.
      When the pRG flow is towards the effective 3p model,  a spontaneous orbital order does not develop, and nematicity is
      associated with
      %due to
     % induced by
      %      a vestigial
      Ising-nematic spin order.
      % caused by stripe SDW fluctuations.
 The phase diagram in Fig.~\ref{fig:full} describes  the behavior found in all four families of FeSCs -- 1111, 122,  111 and 11 systems,
  and in this respect our findings
 provide a unified description of the competition between SDW, SC, and orbital orders in all FeSCs.

We acknowledge with thanks the discussions with  E. Berg, L. Boeri, S. Borisenko, P. Coleman, R. Fernandes, C. Honerkamp, D-H Lee, W. Metzner, A. Nevedomsky, D. Podolsky, M. Scherer, Q. Si, R. Thomale, A-M Tremblay, O. Vafek, C. Varma, and Fa Wang.
L.C. thanks the School of Physics and Astronomy of the University of Minnesota for hospitality during this work and acknowledges funding by the Studienstiftung des deutschen Volkes and the HGSFP at Heidelberg University. A.C. is supported by the Office of Basic Energy Sciences,
 U.S. Department of Energy, under award DE-SC0014402.
MK is supported by the Israel Science Foundation ISF, Grant No. 1287/15 and NSF DMR-1506668.

%\cleardoublepage

%\begin{widetext}
\begin{center}
\textbf{\large Supplemental Material}
\end{center}
%\end{widetext}
\setcounter{equation}{0}
\setcounter{figure}{0}
\setcounter{table}{0}
\setcounter{page}{1}
\makeatletter
\renewcommand{\theequation}{S\arabic{equation}}
\renewcommand{\thefigure}{S\arabic{figure}}
\renewcommand{\bibnumfmt}[1]{[S#1]}
\renewcommand{\citenumfont}[1]{S#1}

%--------------------------------------------------------------------------------------------------------------------------------------------------------------
\section{3-orbital, 5-band model}

%--------------------------------------------------------------------------------------------------------------------------------------------------------------
\subsection{Kinetic part of the Hamiltonian}

We use as an input the fact that the low-energy excitations near all 5 Fermi surfaces are composed  out of three orbitals -- $d_{xz}, d_{yz}$, and $d_{xy}$.
We perform calculations in the 1-Fe unit cell and neglect the dispersion in the third direction and the processes with momentum non-conservation by $(\pi,\pi)$ (the ones which hybridize the pockets).

\begin{figure}[t!]
\centering
 \includegraphics[width=.8\columnwidth]{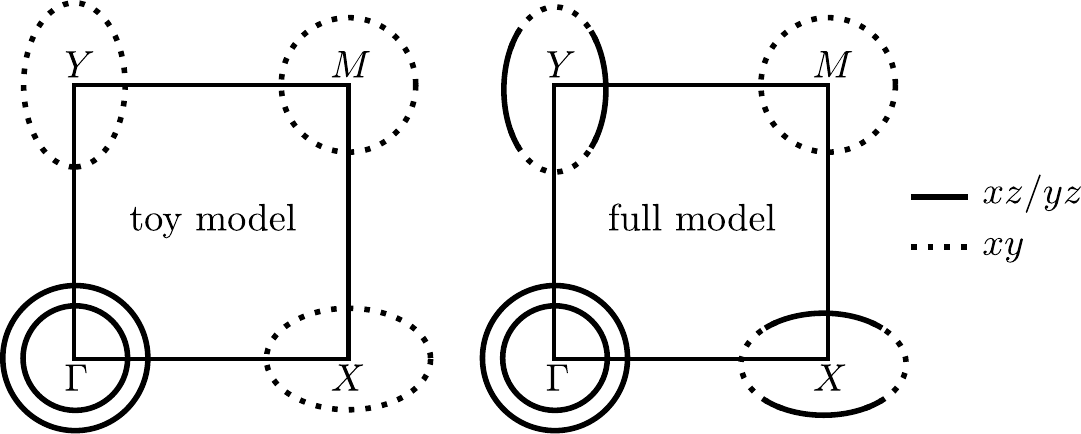}
 \caption{The two  5-pocket models that we consider. The toy and the full model differ in the orbital content of electron pockets. For the full model, the electron pocket at $X$ has contributions from $d_{yz}$ and $d_{xy}$ orbitals and the one at $Y$ has contributions from $d_{xz}$ and $d_{xy}$ orbitals. For the toy model, we approximated these pockets as consisting exclusively of $d_{xy}$ orbital. }
\label{fig:models}
\end{figure}

One way to obtain the dispersion of low-energy excitations is to use the tight-binding model in the orbital basis, restrict with $d_{xz}, d_{yz}$, and $d_{xy}$ orbitals,
and expand around the high-symmetry points in the Brillouin zone, where different electron and hole pockets are located (cf.~Fig.~\ref{fig:models}).
Another way to obtain low-energy dispersions is to identify the symmetry properties around the Fermi level and construct the invariants to leading order in the deviations from the symmetry points\cite{Cvetkovic2013}. The two approaches are equivalent to quadratic order in the deviations near the centra of the pockets ($\Gamma = (0,0)$ for two hole pockets, $M = (\pi,\pi)$ for the third hole pocket, and $X = (\pi,0)$ and $Y=(0,\pi)$ for the two electron pockets).  The effective low-energy Hamiltonian  reads
\begin{equation}
\begin{aligned}
H_0=\sum_{\bf{k},\sigma} &\left[ \psi^\dagger_{\Gamma,\bs{k},\sigma} h_\Gamma(\bs{k}) \psi_{\Gamma,\bs{k},\sigma} + \psi^\dagger_{X,\bs{k},\sigma} h_{X}(\bs{k}) \psi_{X,\bs{k},\sigma}\right. \\
&\left. + \psi^\dagger_{Y,\bs{k},\sigma} h_{Y}(\bs{k}) \psi_{Y,\bs{k},\sigma} + \psi^\dagger_{M,\bs{k},\sigma} h_M(\bs{k}) \psi_{M,\bs{k},\sigma}\right],
\label{1}
\end{aligned}
\end{equation}
where
\small
\begin{align}
h_\Gamma(\bs{k})&\!=\!\begin{pmatrix} \! \epsilon_\Gamma + \frac{k^2}{2m_\Gamma}+ak^2\cos 2\theta_k & ck\sin 2\theta_k \\ ck\sin 2\theta_k & \epsilon_\Gamma + \frac{k^2}{2m_\Gamma}+ak^2\cos 2\theta_k\!\end{pmatrix} \notag\\
h_{X/Y}(\bs{k})&\!=\!\begin{pmatrix}\! \epsilon_1 + \frac{k^2}{2m_1}\pm a_1k^2\cos 2\theta_k & -i v_{X/Y}(\bs{k}) \\ i v_{X/Y}(\bs{k}) & \epsilon_3 + \frac{k^2}{2m_3}\pm a_3
k^2\cos 2\theta_k\!\end{pmatrix}\notag\\
h_M(\bs{k})&\!=\!\epsilon_M - \frac{k^2}{2m_M}\label{eq:kin}
\end{align}
\normalsize
where $v_{X}(k)=2vk\sin\theta$, $v_{Y}(k)=2vk\cos\theta$ and $\theta_k=\arctan{\frac{k_y}{k_x}}$.
Here and below the term $A/B$ (in, e.g., $h_{X/Y}$) means "either A or B''.
The spinors in Eq. (\ref{1})  are defined as $\psi_{\Gamma,\bs{k},\sigma}=(d_{yz,\bs{k},\sigma},d_{xz,\bs{k},\sigma})^T$, $\psi_{X,\bs{k},\sigma}=(d_{yz,\bs{X+k},\sigma},d_{xy,\bs{X+k},\sigma})^T$,$\psi_{Y,\bs{k},\sigma}=(d_{xz,\bs{Y+k},\sigma},d_{xy,\bs{Y+k},\sigma})^T$ and $\psi_{M,\bs{M+k},\sigma}=d_{xy,\bs{k},\sigma}$.  Below we  shorten notations to
$d_{yz,\bs{k},\sigma} = d_{1,\bs{k},\sigma}, d_{xz,\bs{k},\sigma} = d_{2,\bs{k},\sigma}, d_{yz,\bs{X+k},\sigma} = f_{1,\bs{k},\sigma}, d_{xz,\bs{Y+k},\sigma} = f_{2,\bs{k},\sigma},
d_{xy,\bs{X+k},\sigma} = f_{31,\bs{k},\sigma}, d_{xy,\bs{Y+k},\sigma} = f_{32,\bs{k},\sigma}$, and $d_{xy,\bs{k},\sigma} = d_{3,\bs{k},\sigma}$.
 In these notations,  the spinors are
$\psi_{\Gamma,\bs{k},\sigma}=(d_{1,\bs{k},\sigma},d_{2,\bs{k},\sigma})^T$, $\psi_{X/Y,\bs{k},\sigma}=(f_{1/2,\bs{k},\sigma},f_{31/32,\bs{k},\sigma})^T$ and $\psi_{M,\bs{M+k},\sigma}=d_{3,\bs{k},\sigma}$.

  To make RG analysis more tractable we made several simplifications  in Eq. (\ref{eq:kin}).
 For $\Gamma$-centered hole pockets we set $a=c$. Then the transformation from the orbital to the band basis is given by
\begin{align}
\begin{pmatrix}d_{1,\bs{k},\sigma} \\ d_{2,\bs{k},\sigma}\end{pmatrix}=\begin{pmatrix}\cos\theta_k & \sin\theta_k \\ -\sin\theta_k & \cos \theta_k\end{pmatrix}\begin{pmatrix}c_{\bs{k},\sigma} \\ d_{\bs{k},\sigma}\end{pmatrix},
\end{align}
and the  dispersions of fermions $c_{\bs{k},\sigma}$ and $d_{\bs{k},\sigma}$ are isotropic in ${\bf k}$:
\begin{align}
\epsilon_{c/d,\bs{k},\sigma}=-\frac{k^2}{2 m_c/d}
\end{align}
where $m_{c/d}^{-1}=m_\Gamma^{-1} \pm 2a$. The  two hole Fermi surfaces are obviously circular.
 The fermionic Green's functions  in the orbital representation  are related to $G_{c/d}(i\omega,\bs{k})=(i\omega-\epsilon_{c/d,k}-\mu)^{-1}$ in the band representation
 as
\begin{equation}\label{eq:holeprop}
\begin{aligned}
G_{d_1,d_1}(i\omega,\bs{k})&=G_c(i\omega,\bs{k})\cos^2\theta + G_d(i\omega,\bs{k})\sin^2\theta\\
G_{d_2,d_2}(i\omega,\bs{k})&=G_c(i\omega,\bs{k})\sin^2\theta + G_d(i\omega,\bs{k})\cos^2\theta\\
G_{d_1,d_2}(i\omega,\bs{k})&=G_{d_2,d_1}(i\omega,\bs{k})\\
&=\eck{G_d(i\omega,\bs{k})- G_c(i\omega,\bs{k})}\sin\theta\cos\theta\\
\end{aligned}
\end{equation}
A third hole pocket arises around the $M$-point in the Brillouin zone. Here the transformation from orbital to band basis is trivial, because the spectral weight comes entirely from the $d_{xy}$ orbital. The dispersion is given in Eq.~(\ref{eq:kin}), and the corresponding Green's function is
 $G_M(i\omega,\bs{k})=(i\omega+k^2/(2m_M)-\epsilon_M)^{-1}$. The presence of this hole pocket is material dependent and relatively small changes in the system parameters may sink this pocket below the Fermi level (at least at $k_z=0$, when $k_z$ dispersion is included).
   However, such a pocket is definitely present in, e.g., hole-doped  K$_x$Ba$_{1-x}$Fe$_2$As$_2$ and LiFeAs, which motivates to include it into our model.

For electron pockets, the diagonalization of $h_{X}$ ($h_Y$) gives two bands, of which only one crosses the Fermi level and forms the electron pocket around $X$ ($Y$).
 The electron pockets at $X$ and $Y$ are related by $C_4$ symmetry, i.e.~they map onto each other under a rotation by $\pi/2$. Due to the non-diagonal hybridization $v_{X/Y}(k)$, the transformation from orbital to band basis is not a simple rotation. Nevertheless, it can be expressed through
\small
\begin{align}
\begin{pmatrix}e_{1/2} \\ \bar e_{1/2}\end{pmatrix}=e^{i\phi}\begin{pmatrix}e^{i\phi_1}\cos\varphi_{1/2,\theta} & e^{i\phi_2}\sin\varphi_{1/2,\theta} \\ -e^{-i\phi_2}\sin\varphi_{1/2,\theta} & e^{-i\phi_1}\cos \varphi_{1/2,\theta}\end{pmatrix}\begin{pmatrix}f_{1/2} \\ f_{31/32}\end{pmatrix},
\end{align}
\normalsize
where $e_{1/2} = e_{1,\bs{k},\sigma}, e_{2,\bs{k},\sigma}$ and $\bar e_{1/2}$ are operators for band fermions near the electron pockets, and the functions $\varphi_{1/2,\theta}$ and $\phi_{1/2}$ depend on the system parameters and determine the relative spectral weight of $xz/yz$ and $xy$ orbitals. We set $e_{1/2}$ to describe the
 electrons in the band that crosses the Fermi level. The dispersion of these fermions is
 $\xi_{e1}=k^2_x/(2m_{ex}) + k^2_y/(2m_{ey})-\mu_e,   \xi_{e2}=k^2_x/(2m_{ey}) + k^2_y/(2m_{ex})-\mu_e$. For simplicity we assume $m_{ex} = m_{ey} = m_e$, i.e., set
 $\xi_{e1}=\xi_{e2}= \xi_{e}= k^2/(2m_e) -\mu_e$.  We checked that keeping $m_{ex}$ and $m_{ey}$ different  will not change the pRG equations, once we properly rescale the couplings.

   The electron propagator in orbital representation is expressed in terms of low energy fermions as
\begin{equation}
\begin{aligned}
G_{f_{1/2},f_{1/2}}(i\omega,\bs{k})&=G_{e1/e2}(i\omega,\bs{k})\cos^2\varphi_{1/2,\theta} \\
G_{f_{31/32},f_{31/32}}(i\omega,\bs{k})&=G_{e1/e2}(i\omega,\bs{k})\sin^2\varphi_{1/2,\theta}\\
G_{f_{1/2},f_{31/32}}(i\omega,\bs{k})&=G_{f_{31/32},f_{1/2}}(i\omega,\bs{k})^*\\
=G_{e1/e2}&(i\omega,\bs{k})e^{i(\phi_1-\phi_2)}\cos\varphi_{1/2,\theta}\sin\varphi_{1/2,\theta},
\end{aligned}
\end{equation}
where $G_{e1/e2}(i\omega,\bs{k})=(i\omega-\xi_{e})^{-1}$ ($k$ is counted from $X$ in $G_{e1}$ and from $Y$ in $G_{e2}$).

%-------------------------------------------------------------------------------------------------

\subsection{The toy model}

In the toy model, which we analyze in the main text prior to the full one, we
  approximate the orbital content of the two electron pockets as pure $d_{xy}$. In this case, the electron dispersions are already diagonal in the orbital basis, i.e.
   orbital and band representations are identical.
   Our notation for the electron operators is, in this approximation, $\psi_{X/Y,\bs{k},\sigma}=f_{1/2,\bs{k}}$, where 1/2 just labels the pockets.
    This toy model allows  us to study the impact of the fifth pocket in a transparent way.  Furthermore, we expect that the toy  model already captures a substantial
     portion of the physics of the full model because adiabatically changing the tight-binding parameters of the underlying lattice model, one can move the spectral weight from $d_{xz} (d_{yz})$ to $d_{xy}$ orbital everywhere on the electron pockets.  There are, however, several features of the full model, which are not captured by the
       toy model. These are caused by the interactions which involve both  $xz/yz$ and $xy$-orbital states on the electron pockets.

%--------------------------------------------------------------------------------------------------------------------------------------------------------------
\subsection{Interactions}

\subsubsection{The toy model}

\begin{figure*}[t]
\centering
 \includegraphics[width=1.9\columnwidth]{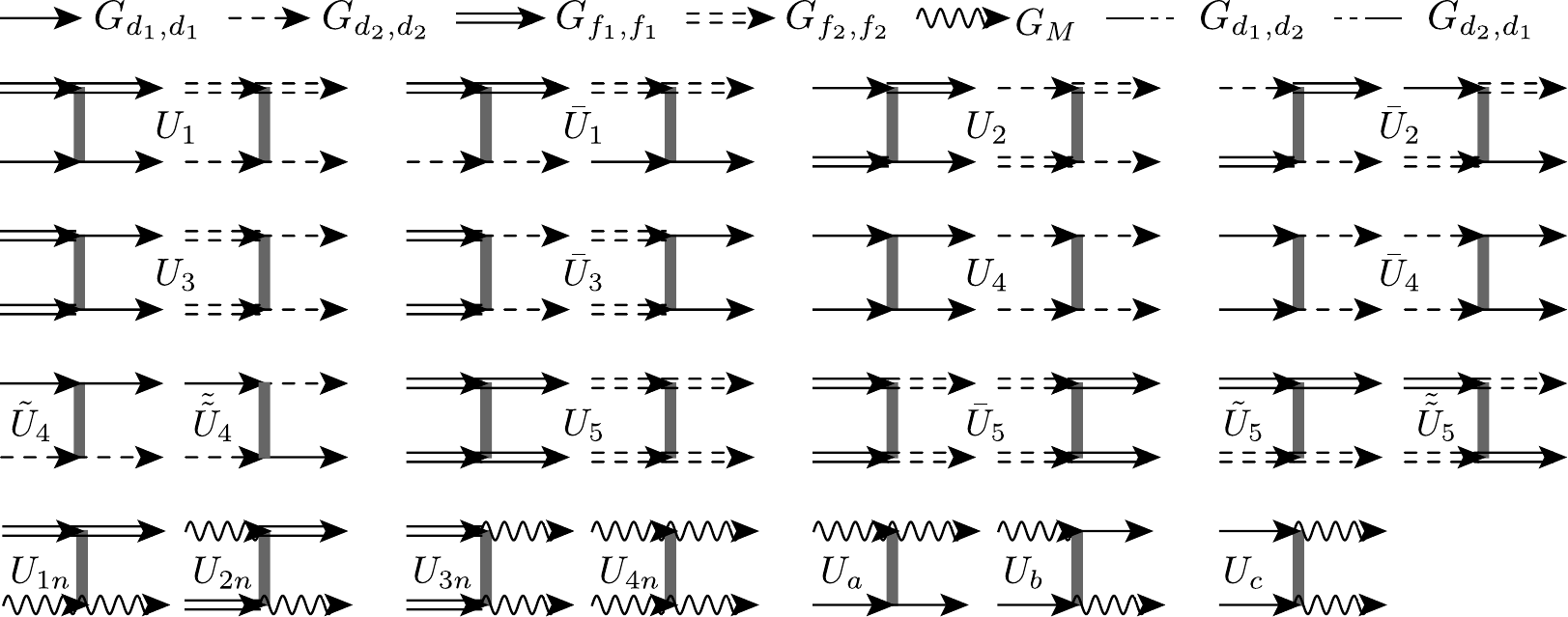}
 \caption{Diagrammatic representation of the 21 interaction terms in the toy model.
 Each interaction term is invariant under $C_4$ rotation.}
\label{fig:iastoy}
\end{figure*}

As we said in the main text, the total number of different interactions between low-energy  fermions in the toy model is 21.
Of them 14 interactions  involve fermions near the two $\Gamma$-centered hole pockets and the two electron pockets, and 7 involve fermions near the third hole pocket.
 In terms of the spinor components defined above, the 14 interaction terms are
\begin{align}\label{eq:4p}
H_I^{4ps}&=U_1 \pr{\sum}\eck{f^\dagger_{1\sigma}f_{1\sigma}d^\dagger_{1\pr{\sigma}}d_{1\pr{\sigma}} + f^\dagger_{2\sigma}f_{2\sigma}d^\dagger_{2\pr{\sigma}}d_{2\pr{\sigma}}} \nonumber\\
&+\bar U_1\pr{\sum}\eck{f^\dagger_{2\sigma}f_{2\sigma}d^\dagger_{1\pr{\sigma}}d_{1\pr{\sigma}} + f^\dagger_{1\sigma}f_{1\sigma}d^\dagger_{2\pr{\sigma}}d_{2\pr{\sigma}}} \nonumber\\
&+ U_2 \pr{\sum}\eck{f^\dagger_{1\sigma}d_{1\sigma}d^\dagger_{1\pr{\sigma}}f_{1\pr{\sigma}} + f^\dagger_{2\sigma}d_{2\sigma}d^\dagger_{2\pr{\sigma}}f_{2\pr{\sigma}}} \nonumber\\
&+\bar U_2\pr{\sum}\eck{f^\dagger_{1\sigma}d_{2\sigma}d^\dagger_{2\pr{\sigma}}f_{1\pr{\sigma}} + f^\dagger_{2\sigma}d_{1\sigma}d^\dagger_{1\pr{\sigma}}f_{2\pr{\sigma}}} \nonumber\\
&+ \frac{U_3}{2} \pr{\sum}\eck{f^\dagger_{1\sigma}d_{1\sigma}f^\dagger_{1\pr{\sigma}}d_{1\pr{\sigma}} + f^\dagger_{2\sigma}d_{2\sigma}f^\dagger_{2\pr{\sigma}}d_{2\pr{\sigma}} + h.c.} \nonumber\\
&+\frac{\bar U_3}{2}\pr{\sum}\eck{f^\dagger_{1\sigma}d_{2\sigma}f^\dagger_{1\pr{\sigma}}d_{2\pr{\sigma}} + f^\dagger_{2\sigma}d_{1\sigma}f^\dagger_{2\pr{\sigma}}d_{1\pr{\sigma}} + h.c.} \nonumber\\
&+\frac{U_4}{2} \pr{\sum}\eck{d^\dagger_{1\sigma}d_{1\sigma}d^\dagger_{1\pr{\sigma}}d_{1\pr{\sigma}} + d^\dagger_{2\sigma}d_{2\sigma}d^\dagger_{2\pr{\sigma}}d_{2\pr{\sigma}}} \nonumber\\
&+\frac{\bar U_4}{2}\pr{\sum}\eck{d^\dagger_{1\sigma}d_{2\sigma}d^\dagger_{1\pr{\sigma}}d_{2\pr{\sigma}} + d^\dagger_{2\sigma}d_{1\sigma}d^\dagger_{2\pr{\sigma}}d_{1\pr{\sigma}}} \nonumber\\
&+ \tilde U_4 \pr{\sum}d^\dagger_{1\sigma}d_{1\sigma}d^\dagger_{2\pr{\sigma}}d_{2\pr{\sigma}} +\tilde{\tilde U}_4\pr{\sum} d^\dagger_{1\sigma}d_{2\sigma}d^\dagger_{2\pr{\sigma}}d_{1\pr{\sigma}} \nonumber\\
&+ \frac{U_5}{2} \pr{\sum}\eck{f^\dagger_{1\sigma}f_{1\sigma}f^\dagger_{1\pr{\sigma}}f_{1\pr{\sigma}} + f^\dagger_{2\sigma}f_{2\sigma}f^\dagger_{2\pr{\sigma}}f_{2\pr{\sigma}}} \nonumber\\
&+\frac{\bar U_5}{2}\pr{\sum}\eck{f^\dagger_{1\sigma}f_{2\sigma}f^\dagger_{1\pr{\sigma}}f_{2\pr{\sigma}} + f^\dagger_{2\sigma}f_{1\sigma}f^\dagger_{2\pr{\sigma}}f_{1\pr{\sigma}}} \nonumber\\
&+ \tilde U_5 \pr{\sum}f^\dagger_{1\sigma}f_{1\sigma}f^\dagger_{2\pr{\sigma}}f_{2\pr{\sigma}} +\tilde{\tilde U}_5\pr{\sum} f^\dagger_{1\sigma}f_{2\sigma}f^\dagger_{2\pr{\sigma}}f_{1\pr{\sigma}},
\end{align}
  where the sum $\pr{\sum}$ denotes the summation over spin $\sigma,\pr{\sigma}$, momenta $\bs{k}_1+\bs{k}_2-\bs{k}_3-\bs{k}_4=0$ and includes the normalization factor $1/N$. The other 7 couplings are
\begin{equation}\label{eq:5p}
\begin{aligned}
H_I^{5p}&=U_{1n} \pr{\sum}\eck{d^\dagger_{3\sigma}d_{3\sigma}f^\dagger_{1\pr{\sigma}}f_{1\pr{\sigma}} + d^\dagger_{3\sigma}d_{3\sigma}f^\dagger_{2\pr{\sigma}}f_{2\pr{\sigma}}} \\
&+U_{2n} \pr{\sum}\eck{d^\dagger_{3\sigma}f_{1\sigma}f^\dagger_{1\pr{\sigma}}d_{3\pr{\sigma}} + d^\dagger_{3\sigma}f_{2\sigma}f^\dagger_{2\pr{\sigma}}d_{3\pr{\sigma}}} \\
&+\frac{U_{3n}}{2} \pr{\sum}\eck{d^\dagger_{3\sigma}f_{1\sigma}d^\dagger_{3\pr{\sigma}}f_{1\pr{\sigma}} + d^\dagger_{3\sigma}f_{2\sigma}d^\dagger_{3\pr{\sigma}}f_{2\pr{\sigma}}+h.c.} \\
&+\frac{U_{4n}}{2} \pr{\sum}d^\dagger_{3\sigma}d_{3\sigma}d^\dagger_{3\pr{\sigma}}d_{3\pr{\sigma}} \\
&+U_a \pr{\sum}\eck{d^\dagger_{3\sigma}d_{3\sigma}d^\dagger_{1\pr{\sigma}}d_{1\pr{\sigma}} + d^\dagger_{3\sigma}d_{3\sigma}d^\dagger_{2\pr{\sigma}}d_{2\pr{\sigma}}} \\
&+U_b \pr{\sum}\eck{d^\dagger_{3\sigma}d_{1\sigma}d^\dagger_{1\pr{\sigma}}d_{3\pr{\sigma}} + d^\dagger_{3\sigma}d_{2\sigma}d^\dagger_{2\pr{\sigma}}d_{3\pr{\sigma}}} \\
&+\frac{U_c}{2} \pr{\sum}\eck{d^\dagger_{3\sigma}d_{1\sigma}d^\dagger_{3\pr{\sigma}}d_{1\pr{\sigma}} + d^\dagger_{3\sigma}d_{2\sigma}d^\dagger_{3\pr{\sigma}}d_{2\pr{\sigma}}+h.c.}
\end{aligned}
\end{equation}
The interactions of the toy model are sketched in Fig.~\ref{fig:iastoy}. Each single interaction term in Eq.~(\ref{eq:4p}) and Eq.~(\ref{eq:5p}) obeys the $C_4$ symmetry separately, which is why they do not need to flow equally under RG.

The bare values of the 21 couplings are expressed in terms of the parameters of the microscopic model for intra-orbital and inter-orbital
 interactions between fermions. The commonly used model approximates all interactions as local in real space:
\begin{equation}\label{eq:bareia}
\begin{aligned}
H_I&=U\sum_{i,\mu}n_{i,\mu,\uparrow}n_{i,\mu,\downarrow} + \frac{\pr{U}}{2}\sum_{i,\mu\neq\pr{\mu}}n_{i,\mu}n_{i,\pr{\mu}} \\
&+ \frac{J}{2}\sum_{i,\mu\neq\pr{\mu}}\sum_{\sigma,\pr{\sigma}}d^\dagger_{i,\mu,\sigma}d^\dagger_{i,\pr{\mu},\pr{\sigma}}d_{i,\pr{\mu},\pr{\sigma}}d_{i,\mu,\sigma} \\
&+ \frac{\pr{J}}{2}\sum_{i,\mu\neq\pr{\mu}}\sum_{\sigma,\pr{\sigma}}d^\dagger_{i,\mu,\sigma}d^\dagger_{i,\mu,\pr{\sigma}}d_{i,\pr{\mu},\pr{\sigma}}d_{i,\pr{\mu},\sigma}.
\end{aligned}
\end{equation}
Here the sums run over the sites $i$, the spin components $\sigma$, and the three orbitals $\mu=xy,xz,yz$. The density operator on site $i$ in orbital $\mu$ is labeled by $n_{i,\mu}=\sum_{\sigma}n_{i,\mu,\sigma}$ and $n_{i,\mu,\sigma}=d^\dagger_{i,\mu,\sigma} d_{i,\mu,\sigma}$.
The interactions in Eq. (\ref{eq:bareia})
involve  the Hubbard interaction $U$ between electrons on the same orbital, the onsite repulsion $\pr{U}$ between electrons in different orbitals, the Hund's rule coupling $J$ and the pair-hopping term $\pr{J}$.

By comparing with Eq.~(\ref{eq:bareia}), we obtain the bare values of the 21 couplings
\begin{equation}
\begin{aligned}
U&=U_4=U_5=\bar U_5=\tilde U_5=\tilde{\tilde U}_5=U_{1n}=U_{2n}=U_{3n}=U_{4n}\\
\pr{U}&=U_1=\bar U_1=\tilde U_4=U_a\\
J&=U_2=\bar U_2=\tilde{\tilde U}_4=U_b\\
\pr{J}&=U_3=\bar{U}_3=\bar U_4=U_c
\end{aligned}
\end{equation}

Like we sad in the main text, the 21 interactions all flow to different values under pRG.  This implies that the system self-generates
longer-ranged interactions as one progressively integrates out fermions with higher energies.

\subsubsection{The full model}

\begin{figure*}[t]
\centering
 \includegraphics[width=1.9\columnwidth]{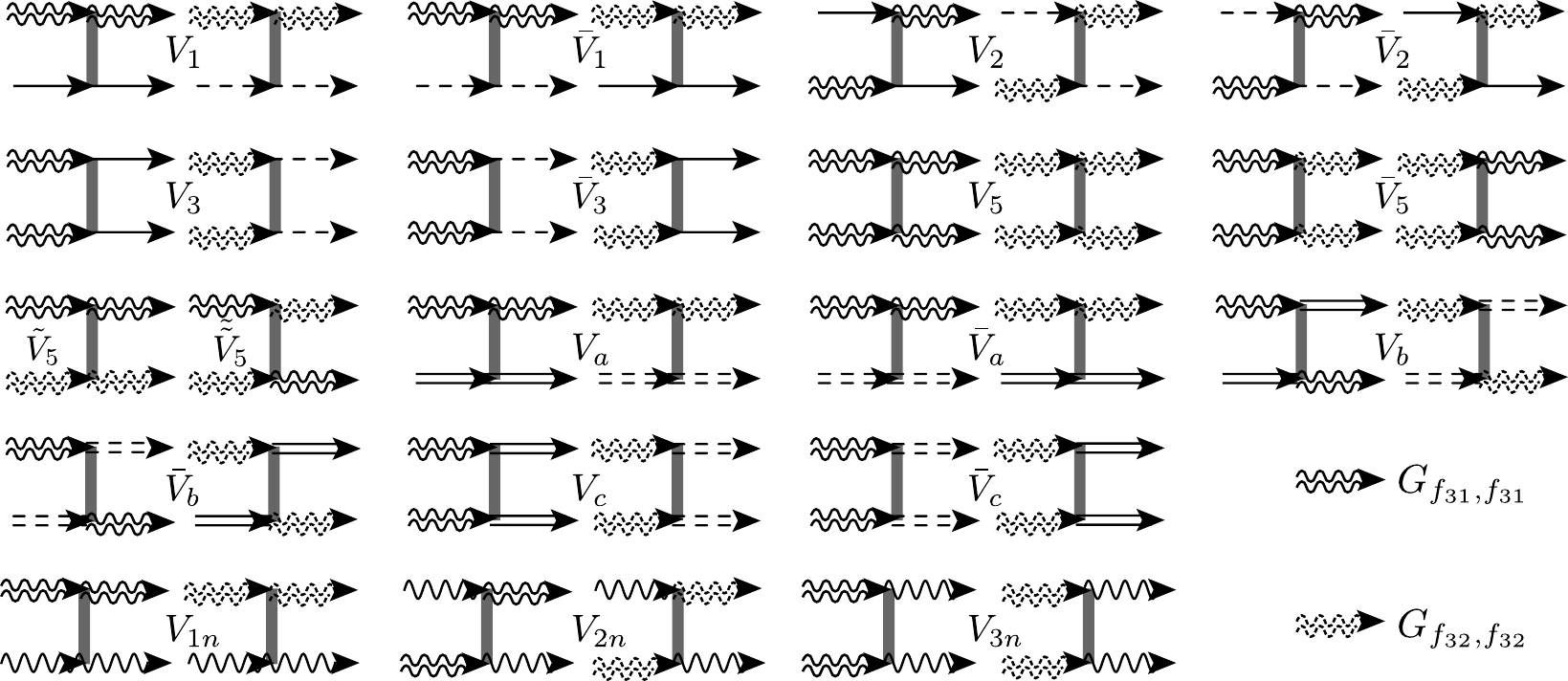}
 \caption{Additional interactions allowed by $C_4$ symmetry in the full model.}
\label{fig:iasfull}
\end{figure*}

In the full model with $d_{xz}/d_{xy}$ and $d_{yz}/d_{xy}$ orbital content of fermions near the electron pockets,
 23 more couplings are allowed by symmetry, what increases the total number of the $C_4$-symmetric interaction terms to 44.
  40 interactions involve  pairs of fermions, with each pair near either $\Gamma, X, Y$, or $M$ point.  The 4 additional interactions involve fermions one near each of these  points. 
  Of the 23 new couplings,
 13  are obtained by substituting $f_1$ and $f_2$ by $f_{31}$ and $f_{32}$ in  Eqs.~(\ref{eq:4p}) and (\ref{eq:5p}):
\begin{align}\label{eq:iafull1}
H_I^{(1)}&=V_1 \pr{\sum}\eck{f^\dagger_{31\sigma}f_{31\sigma}d^\dagger_{1\pr{\sigma}}d_{1\pr{\sigma}} + f^\dagger_{32\sigma}f_{32\sigma}d^\dagger_{2\pr{\sigma}}d_{2\pr{\sigma}}} \nonumber\\
&+\bar V_1\pr{\sum}\eck{f^\dagger_{32\sigma}f_{32\sigma}d^\dagger_{1\pr{\sigma}}d_{1\pr{\sigma}} + f^\dagger_{31\sigma}f_{31\sigma}d^\dagger_{2\pr{\sigma}}d_{2\pr{\sigma}}} \nonumber\\
&+ V_2 \pr{\sum}\eck{f^\dagger_{31\sigma}d_{1\sigma}d^\dagger_{1\pr{\sigma}}f_{31\pr{\sigma}} + f^\dagger_{32\sigma}d_{2\sigma}d^\dagger_{2\pr{\sigma}}f_{32\pr{\sigma}}} \nonumber\\
&+\bar V_2\pr{\sum}\eck{f^\dagger_{31\sigma}d_{2\sigma}d^\dagger_{2\pr{\sigma}}f_{31\pr{\sigma}} + f^\dagger_{32\sigma}d_{1\sigma}d^\dagger_{1\pr{\sigma}}f_{32\pr{\sigma}}} \nonumber\\
&+ \frac{V_3}{2} \pr{\sum}\eck{f^\dagger_{31\sigma}d_{1\sigma}f^\dagger_{31\pr{\sigma}}d_{1\pr{\sigma}} + f^\dagger_{32\sigma}d_{2\sigma}f^\dagger_{32\pr{\sigma}}d_{2\pr{\sigma}} + h.c.} \nonumber\\
&+\frac{\bar V_3}{2}\pr{\sum}\eck{f^\dagger_{31\sigma}d_{2\sigma}f^\dagger_{31\pr{\sigma}}d_{2\pr{\sigma}} + f^\dagger_{32\sigma}d_{1\sigma}f^\dagger_{32\pr{\sigma}}d_{1\pr{\sigma}} + h.c.} \nonumber\\
&+ \frac{V_5}{2} \pr{\sum}\eck{f^\dagger_{31\sigma}f_{31\sigma}f^\dagger_{31\pr{\sigma}}f_{31\pr{\sigma}} + f^\dagger_{32\sigma}f_{32\sigma}f^\dagger_{32\pr{\sigma}}f_{32\pr{\sigma}}} \nonumber\\
&+\frac{\bar V_5}{2}\pr{\sum}\eck{f^\dagger_{31\sigma}f_{32\sigma}f^\dagger_{31\pr{\sigma}}f_{32\pr{\sigma}} + f^\dagger_{32\sigma}f_{31\sigma}f^\dagger_{32\pr{\sigma}}f_{31\pr{\sigma}}}\nonumber \\
&+ \tilde V_5 \pr{\sum}f^\dagger_{31\sigma}f_{31\sigma}f^\dagger_{32\pr{\sigma}}f_{32\pr{\sigma}} +\tilde{\tilde V}_5\pr{\sum} f^\dagger_{31\sigma}f_{32\sigma}f^\dagger_{32\pr{\sigma}}f_{31\pr{\sigma}} \nonumber\\
&+V_{1n} \pr{\sum}\eck{d^\dagger_{3\sigma}d_{3\sigma}f^\dagger_{31\pr{\sigma}}f_{31\pr{\sigma}} + d^\dagger_{3\sigma}d_{3\sigma}f^\dagger_{32\pr{\sigma}}f_{32\pr{\sigma}}} \nonumber\\
&+V_{2n} \pr{\sum}\eck{d^\dagger_{3\sigma}f_{31\sigma}f^\dagger_{31\pr{\sigma}}d_{3\pr{\sigma}} + d^\dagger_{3\sigma}f_{32\sigma}f^\dagger_{32\pr{\sigma}}d_{3\pr{\sigma}}} \nonumber\\
&+\frac{V_{3n}}{2} \pr{\sum}\eck{d^\dagger_{3\sigma}f_{31\sigma}d^\dagger_{3\pr{\sigma}}f_{31\pr{\sigma}} + d^\dagger_{3\sigma}f_{32\sigma}d^\dagger_{3\pr{\sigma}}f_{32\pr{\sigma}}+h.c.}
\end{align}
Further six couplings come from interactions involving $xy$ and $xz/yz$ orbital states on the electron pockets
\begin{align}\label{eq:iafull12}
H_I^{(2)}&=V_a \pr{\sum}\eck{f^\dagger_{31\sigma}f_{31\sigma}f^\dagger_{1\pr{\sigma}}f_{1\pr{\sigma}} + f^\dagger_{32\sigma}f_{32\sigma}f^\dagger_{2\pr{\sigma}}f_{2\pr{\sigma}}} \nonumber\\
&+\bar V_a\pr{\sum}\eck{f^\dagger_{32\sigma}f_{32\sigma}f^\dagger_{1\pr{\sigma}}f_{1\pr{\sigma}} + f^\dagger_{31\sigma}f_{31\sigma}f^\dagger_{2\pr{\sigma}}f_{2\pr{\sigma}}} \nonumber\\
&+ V_b \pr{\sum}\eck{f^\dagger_{31\sigma}f_{1\sigma}f^\dagger_{1\pr{\sigma}}f_{31\pr{\sigma}} + f^\dagger_{32\sigma}f_{2\sigma}f^\dagger_{2\pr{\sigma}}f_{32\pr{\sigma}}} \nonumber\\
&+\bar V_b\pr{\sum}\eck{f^\dagger_{31\sigma}f_{2\sigma}f^\dagger_{2\pr{\sigma}}f_{31\pr{\sigma}} + f^\dagger_{32\sigma}f_{1\sigma}f^\dagger_{1\pr{\sigma}}f_{32\pr{\sigma}}} \nonumber\\
&+ \frac{V_c}{2} \pr{\sum}\eck{f^\dagger_{31\sigma}f_{1\sigma}f^\dagger_{31\pr{\sigma}}f_{1\pr{\sigma}} + f^\dagger_{32\sigma}f_{2\sigma}f^\dagger_{32\pr{\sigma}}f_{2\pr{\sigma}} + h.c.} \nonumber\\
&+\frac{\bar V_c}{2}\pr{\sum}\eck{f^\dagger_{31\sigma}f_{2\sigma}f^\dagger_{31\pr{\sigma}}f_{2\pr{\sigma}} + f^\dagger_{32\sigma}f_{1\sigma}f^\dagger_{32\pr{\sigma}}f_{1\pr{\sigma}} + h.c.}
\end{align}
We show these
 interactions graphically in
 Fig.~\ref{fig:iasfull}. Note that, in contrast to the simplified model, $f_{1/2}$ now labels fermions with $yz/xz$ orbital content, whereas $f_{31,32}$ labels fermions with  $xy$ orbital content.

 Finally there are four additional interactions that, in contrast to the previous 40 interactions, involve fermions near each of the four high-symmetry points $\Gamma,X,Y,M$. 
 In explicit form, these interactions are
 \begin{align}
H_I^{(3)}&=W_1 \pr{\sum}\eck{f^\dagger_{1\sigma}d_{3\sigma}f^\dagger_{32\pr{\sigma}}d_{1\pr{\sigma}} + f^\dagger_{2\sigma}d_{3\sigma}f^\dagger_{31\pr{\sigma}}d_{2\pr{\sigma}}+h.c.} \nonumber\\
&+W_2 \pr{\sum}\eck{f^\dagger_{31\sigma}d_{3\sigma}f^\dagger_{2\pr{\sigma}}d_{2\pr{\sigma}} + f^\dagger_{32\sigma}d_{3\sigma}f^\dagger_{1\pr{\sigma}}d_{1\pr{\sigma}}+h.c.} \nonumber\\
&+W_3 \pr{\sum}\eck{f^\dagger_{1\sigma}d_{1\sigma}d^\dagger_{3\pr{\sigma}}f_{32\pr{\sigma}} + f^\dagger_{2\sigma}d_{2\sigma}d^\dagger_{3\pr{\sigma}}f_{31\pr{\sigma}}+h.c.} \nonumber\\
&+W_4 \pr{\sum}\eck{f^\dagger_{31\sigma}d_{2\sigma}d^\dagger_{3\pr{\sigma}}f_{2\pr{\sigma}} + f^\dagger_{32\sigma}d_{1\sigma}d^\dagger_{3\pr{\sigma}}f_{1\pr{\sigma}}+h.c.}
\end{align}
We checked explicitly that these four additional interactions
 do not affect the behavior near each of the four stable fixed trajectories, which we obtained by solving the
  pRG equations for 40 couplings (see Sec.~\ref{sec:fullprg}).  This is what we presented in the main text.  We also verified that these additional interactions do not  generate new fixed trajectories, if the bare values of these interactions are within certain limits.  Outside these limits, the 4 additional interactions
   may, in principle,
   move the system towards a new stable fixed trajectory.  We did not explore this possibility here and in the following  we neglect these four additional interactions.

 Like we did for the toy model, we express the bare values of the 40 couplings in terms of $U$, $U'$, $J$, $J'$. We have
 % and are given by
\begin{equation}\label{eq:barefull}
\begin{aligned}
U&=U_1=U_2=U_3=U_4=U_5=\bar U_5=U_{4n}=V_5=\bar V_5\\
&=\tilde V_5 =\tilde{\tilde V}_5=V_{1n}=V_{2n}=V_{3n}\\
\pr{U}&=\bar U_1=\tilde U_4=\tilde U_5=U_{1n}=U_a=V_a=\bar V_a=V_1=\bar V_1\\
J&=\bar U_2=\tilde{\tilde U}_4=\tilde{\tilde U}_5=U_{2n}=U_b=V_b=\bar V_b=V_2=\bar V_2\\
\pr{J}&=\bar{U}_3=\bar U_4=\bar U_5=U_{3n}=U_c=V_c=\bar V_c=V_3=\bar V_3.
\end{aligned}
\end{equation}

%--------------------------------------------------------------------------------------------------------------------------------------------------------------
\section{Analytic parguet RG for 5-pocket model}

We employ a pRG approach to study the hierarchy of the  orders that the system develops at low energies. The pRG procedure allows us to see how the susceptibilities in
 different ordering channels evolve as the system flows to low energies, including their mutual feedback. In the pRG procedure, one integrates out fermions
  with energies  down to a
   progressively smaller running energy $E$ and observes how the couplings vary as $E$ gets smaller.
   We describe this flow of interactions in terms of the RG  scale $L=\log{\Lambda/E}$, where $\Lambda$ is the UV-cutoff, generally of the order of the bandwidth.
     The logarithmic energy scale $L$ appears due to the fact that the polarization bubbles in the particle-particle channel at zero total momentum
      {\it and} the particle-hole channel at momenta $(\pi,0)$ and $(0,\pi)$ are logarithmical.
  As a result of the integration procedure, we obtain coupled differential equations -the flow equations- for all the interactions, describing their evolution with $L$.
    We solve for the running couplings $U_i (L)$ and use these solutions as inputs  to calculate susceptibilities in different ordering channels (SDW, CDW, superconducting and Pomeranchul channels). An instability in a particular channel is signaled by the divergence of the corresponding susceptibility at a scale $L_{cr}$.
     Below we show the details of pRG analysis for the toy model and the full model.
      We recall that pRG analysis works when $E$ is larger than the Fermi energy, i.e., when $L < L_F = \log{\Lambda/E_F}$ (see, e.g., Ref. \cite{maiti2010}).
         If $L_{cr} < L_F$, the pRG analysis works all the way to the leading instability.
         If $L_{cr} > L_F$, pRG analysis allows one to determine the largest susceptibility at $L = L_F$.  It is likely (although not guaranteed) that this susceptibility will diverge first at a lower energy.

%--------------------------------------------------------------------------------------------------------------------------------------------------------------
\subsection{PRG for the toy model}
\subsubsection{PRG equations and fixed trajectories}

\begin{figure*}[t]
\centering
 \includegraphics[width=.85\columnwidth]{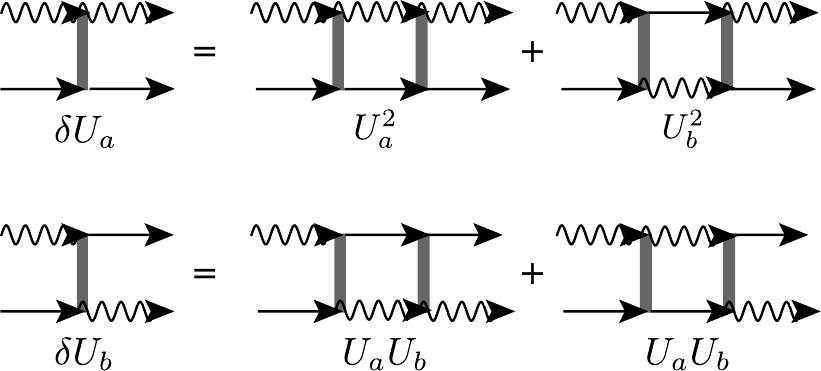}
 \caption{Diagrammatic representation of the 1-loop renormalizations of the interactions $U_a$ and $U_b$. They decouple from the remaining interactions and are representative for the subgroup of interactions flowing to zero.}
\label{fig:loopsUaUb}
\end{figure*}

\begin{figure*}[t]
\centering
 \includegraphics[width=1.85\columnwidth]{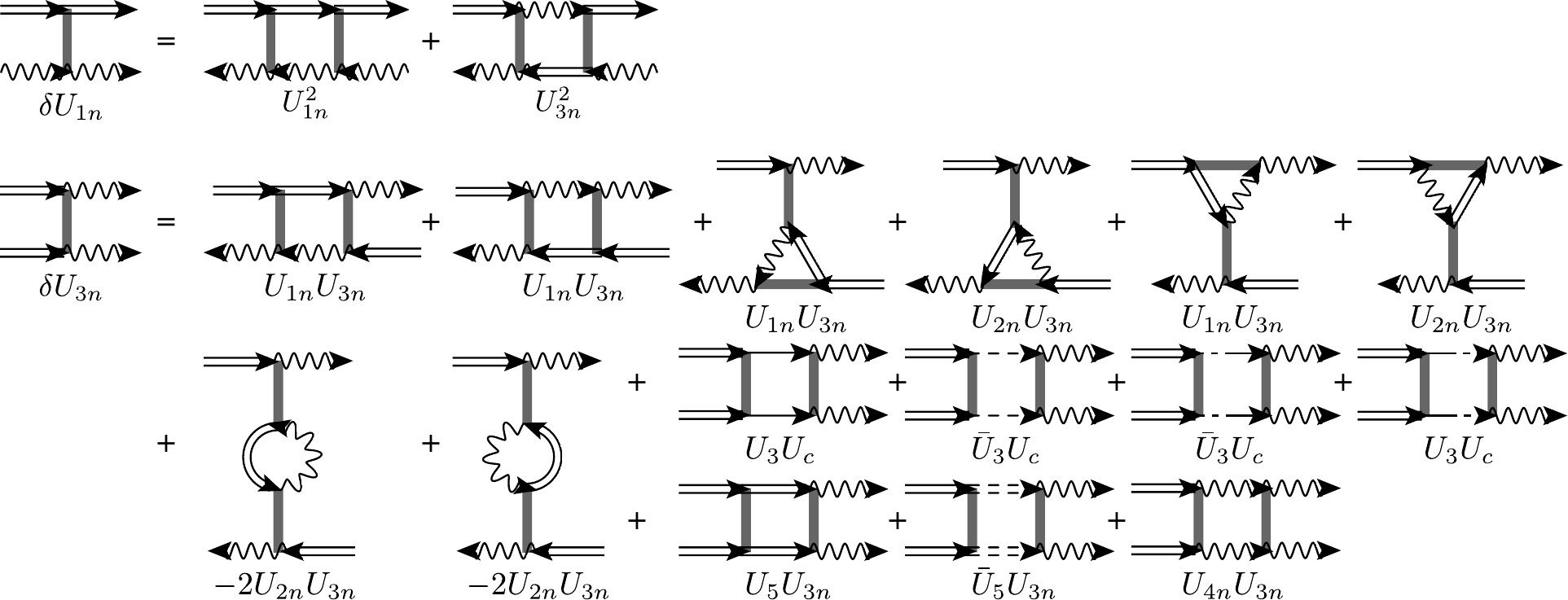}
 \caption{Diagrammatic representation of the 1-loop renormalization of the interactions $U_{1n}$ and $U_{3n}$.}
\label{fig:loopsU1nU3n}
\end{figure*}

We derive the pRG equations by collecting all possible one-loop diagrams that contribute to logarithmic renormalization of each of the interactions.
  The procedure has been described  Ref.~\cite{we_last} (for a simplified 4-pocket, two-orbital model) and in Ref.~\cite{maiti2010} for 3-pocket, one-orbital model.
   We follow the same line of reasoning as in these works.
   We obtain the pRG equations for our 5-pocket model by combining and modifying  pRG equations from these two models.

     Like in Ref. ~\cite{we_last} we find that pRG equations for 6 combinations of the couplings
     $(\tilde U_4 \pm \tilde{\tilde U}_4),(\tilde U_5 \pm \tilde{\tilde U}_5)$ and $(\tilde U_a \pm \tilde{\tilde U}_b)$ decouple from other RG equations, and these combinations all flow to zero if their bare values are positive, which is the case for $\pr{U}\geq J$.
      We assume that this inequality holds.  If it does not hold, the system may develop a superconducting instability in the spin-triplet $A_{2g}$ channel (Ref. \cite{vafek2016}).
      Representative diagrams for the renormalizations of the couplings from this  group of 6 are shown in Fig.~\ref{fig:loopsUaUb}. The 6 pRG equations
       are:
\begin{equation}\label{eq:irrelevant}
\begin{aligned}
4\pi \frac{d}{dL}\rund{\tilde U_4 \pm \tilde{\tilde U}_4}&=-c^{(1)}_{pp}\rund{\tilde U_4 \pm \tilde{\tilde U}_4}^2\\
4\pi \frac{d}{dL}\rund{\tilde U_5 \pm \tilde{\tilde U}_5}&=-c^{(2)}_{pp}\rund{\tilde U_5 \pm \tilde{\tilde U}_5}^2\\
4\pi \frac{d}{dL}\rund{ U_a \pm U_b}&=-c^{(3)}_{pp}\rund{U_a \pm U_b}^2,
\end{aligned}
\end{equation}
where
 $c_{pp}^{(1)}=\frac{1}{8}(m_c+m_d+12 \frac{m_c m_d}{m_c+m_d}\pm\frac{(m_c-m_d)^2}{m_c+m_d})$, $c_{pp}^{(2)}=m_e$ and $c_{pp}^{(3)}=\frac{m_M m_c}{m_M+m_c}+\frac{m_M m_d}{m_M+m_d}$.
The pRG equations for the other remaining 15 couplings are
\begin{align}\label{eq:rgxy}
4\pi \frac{d}{dL} U_1&=A(U_1^2+U_3^2)\nonumber\\
4\pi \frac{d}{dL} \bar U_1&=A(\bar U_1^2+\bar U_3^2)\nonumber\\
4\pi \frac{d}{dL} U_{1n}&=A_n(U_{1n}^2+U_{3n}^2)\nonumber\\
4\pi \frac{d}{dL} U_2&=2AU_2(U_1-U_2)\nonumber\\
4\pi \frac{d}{dL} \bar U_2&=2A\bar U_2(\bar U_1-\bar U_2)\nonumber\\
4\pi \frac{d}{dL} U_{2n}&=2A_nU_{2n}(U_{1n}-U_{2n})\nonumber\\
4\pi \frac{d}{dL} U_3&=2AU_3(2U_1-U_2)-A_e(U_3U_5+\bar U_3\bar U_5) \nonumber\\
-A_h &(U_3U_{4}+\bar U_3\bar U_{4})-A_h^-(U_3\bar U_4 +\bar U_3 U_4)-A_M U_{3n}U_c\nonumber\\
4\pi \frac{d}{dL} \bar U_3&=2A\bar U_3(2\bar U_1-\bar U_2)-A_e(\bar U_3U_5+ U_3\bar U_5) \nonumber\\
-A_h &(\bar U_3U_{4}+ U_3\bar U_{4})-A_h^-(U_3U_4 +\bar U_3 \bar U_4)-A_M U_{3n}U_c\nonumber\\
4\pi \frac{d}{dL} U_{3n}&=2A_nU_{3n}(2U_{1n}-U_{2n})-A_eU_{3n}(U_5+\bar U_5)\nonumber\\
-A_M&U_{3n}U_{4n}-(A_h+A_h^-)(U_3+\bar U_3) U_c\nonumber\\
4\pi \frac{d}{dL} U_4&=-A_h(U_4^2+\bar U_4^2)-2A_h^-U_4\bar U_4-A_e(U_3^2+\bar U_3^2)\nonumber\\
&-A_MU_c^2\nonumber\\
4\pi \frac{d}{dL} \bar U_4&=-2A_hU_4\bar U_4-A_h^-(U_4^2+\bar U_4^2)-2A_eU_3\bar U_3\nonumber\\
&-A_MU_c^2\nonumber\\
4\pi \frac{d}{dL} U_{4n}&=-A_MU_{4n}^2-2A_eU_{3n}^2-2(A_h+A_h^-)U_c^2\nonumber\\
4\pi \frac{d}{dL} U_5&=-A_e(U_5^2+\bar U_5^2)-A_h(U_3^2+\bar U_3^2)-2A_h^-U_3\bar U_3\nonumber\\
&-A_MU_{3n}^2\nonumber\\
4\pi \frac{d}{dL} \bar U_5&=-2A_eU_5\bar U_5-2A_hU_3\bar U_3-A_h^-(U_3^2+\bar U_3^2)\nonumber\\
&-A_MU_{3n}^2\nonumber\\
4\pi \frac{d}{dL} U_c&=-(A_h+A_h^-)(U_4+\bar U_4)U_c-A_MU_{4n}U_c\nonumber\\
&-A_3(U_3+\bar U_3)U_{3n}.
\end{align}
As an example, the one-loop diagrams that renormalize $U_{1n}$ and $U_{3n}$ are presented in Fig.~\ref{fig:loopsU1nU3n}.
The numerical prefactors in the r.h.s. of pRG equations are
 $A=\frac{m_e m_c}{m_e+m_c}+\frac{m_e m_d}{m_e+m_d}$, $A_n=\frac{m_Mm_e}{m_M+m_e}$, $A_e=m_e$, $A_h=\frac{3}{8}(m_c+m_d)+\frac{1}{2}\frac{m_cm_d}{m_c+m_d}$, $A_h^-=\frac{1}{8}\frac{(m_c-m_d)^2}{m_c+m_d}$ and $A_M=m_M$.
Note that the contribution $A_h^-=(m_c-m_d)^2/(8(m_c+m_d))$ comes from the $G_{d_1,d_2}$  the propagator for fermions near the $\Gamma-$centered hole pockets (see Eq.(\ref{eq:holeprop})).

   To proceed, we note that, if $U_i=\bar U_i$,  then $d_LU_i=d_L\bar U_i$. We have checked that the trajectory with this property is a stable one.
    We searched for other potential stable fixed trajectories, but did not find one.  Hence we set $U_i=\bar U_i$.  We further introduce the dimensionless couplings
 $u_{1,2}=A/(4\pi) U_{1,2}$, $u_3=A/(4\pi) aU_3$, $u_4=A_h/(4\pi)U_4$, $u_5=A_e/(4\pi)U_5$, $u_{1n,2n}=A_n/(4\pi)U_{1n,2n}$, $u_{3n}=A_n/(4\pi) a_nU_{3n}$, $u_{4n}=A_M/(4\pi)U_{4n}$ and $u_{5n}=\sqrt{A_MA_h}/(4\pi)U_c$ and define $a=\sqrt{A_hA_e}/A$ and $a_n=\sqrt{A_MA_e}/A_n$ and $b=1+A_h^-/A_h$. Then we obtain the pRG equations
\begin{align}
&\dot u_1 =u_1^2+\frac{u_3^2}{a^2} \\
&\dot u_{1n}=u_{1n}^2+\frac{u_{3n}^2}{a_n^2} \nonumber\\
&\dot u_2 =2u_2(u_1-u_2)\nonumber\\
&\dot u_{2n}=2u_{2n}(u_{1n}-u_{2n}) \nonumber \\
& \dot u_3=2u_3(2u_1-u_2-u_5)-2 b u_3u_{4}-u_{3n}u_{5n} \nonumber\\
&\dot u_{3n}=2u_{3n}(2u_{1n}-u_{2n}-u_5)-u_{3n}u_{4n}-2 b u_3 u_{5n} \nonumber \\
&\dot u_4 =-2b u_4^2-2u_3^2-2u^2_{5n}\nonumber\\
&\dot u_{4n}=-u_{4n}^2-2u_{3n}^2-2b u^2_{5n} \nonumber\\
&\dot u_5=-2u_5^2-2b u_3^2-u_{3n}^2\nonumber\\
&\dot u_{5n}=-2bu_4u_{5n}-u_{4n}u_{5n}-2u_3u_{3n},\nonumber
\end{align}
 which we presented in the main text for $a=a_n=b=1$.

\subsubsection{The solution of PRG equations}

  To simplify the analysis  we  assume $m_c\approx m_d$ and neglect the contribution from $A_h^-$, i.e. set $A_h^-=0$.
\begin{figure}[t!]
\centering
 \includegraphics[width=.9\columnwidth]{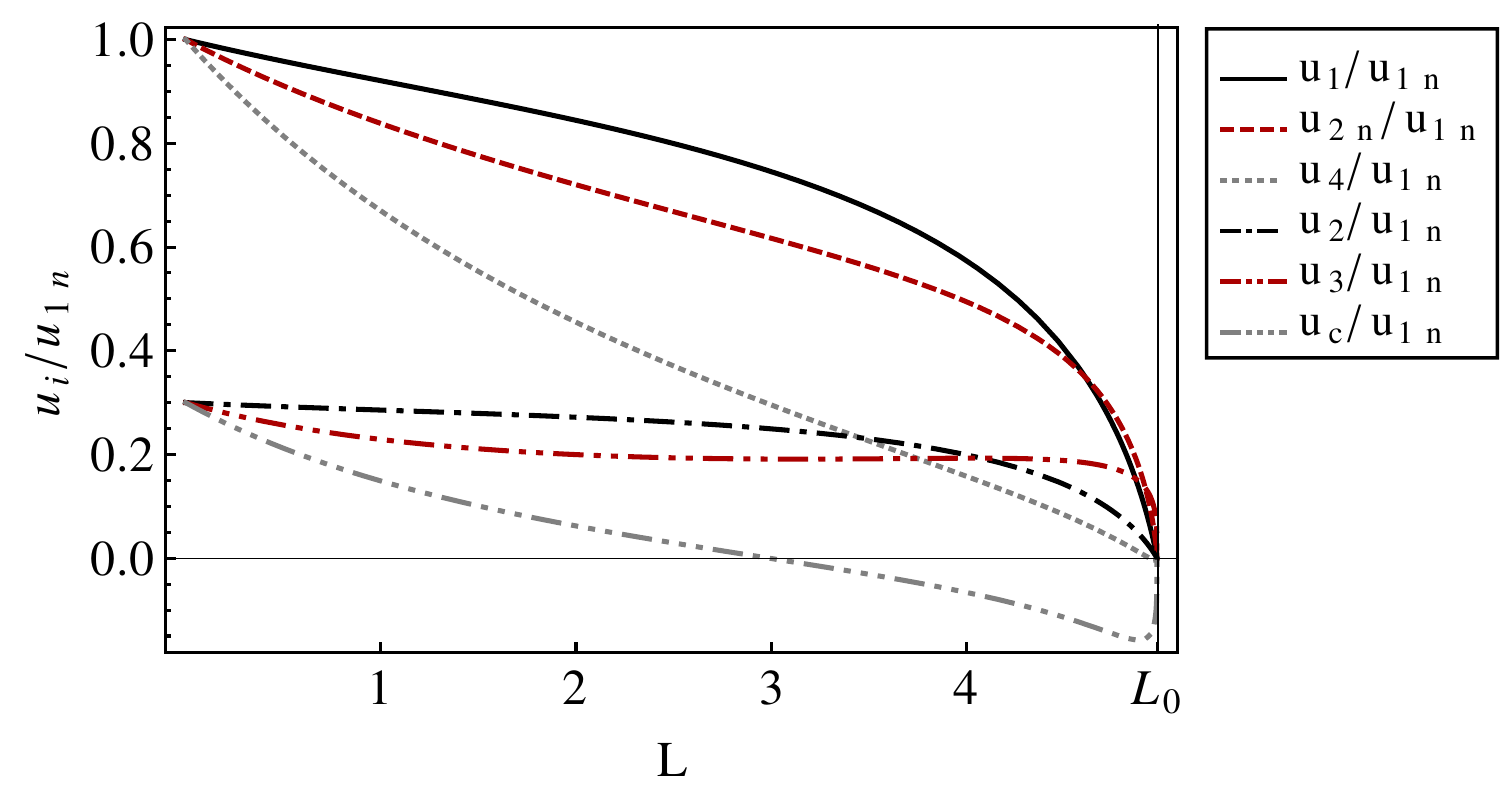}\\
  \includegraphics[width=.9\columnwidth]{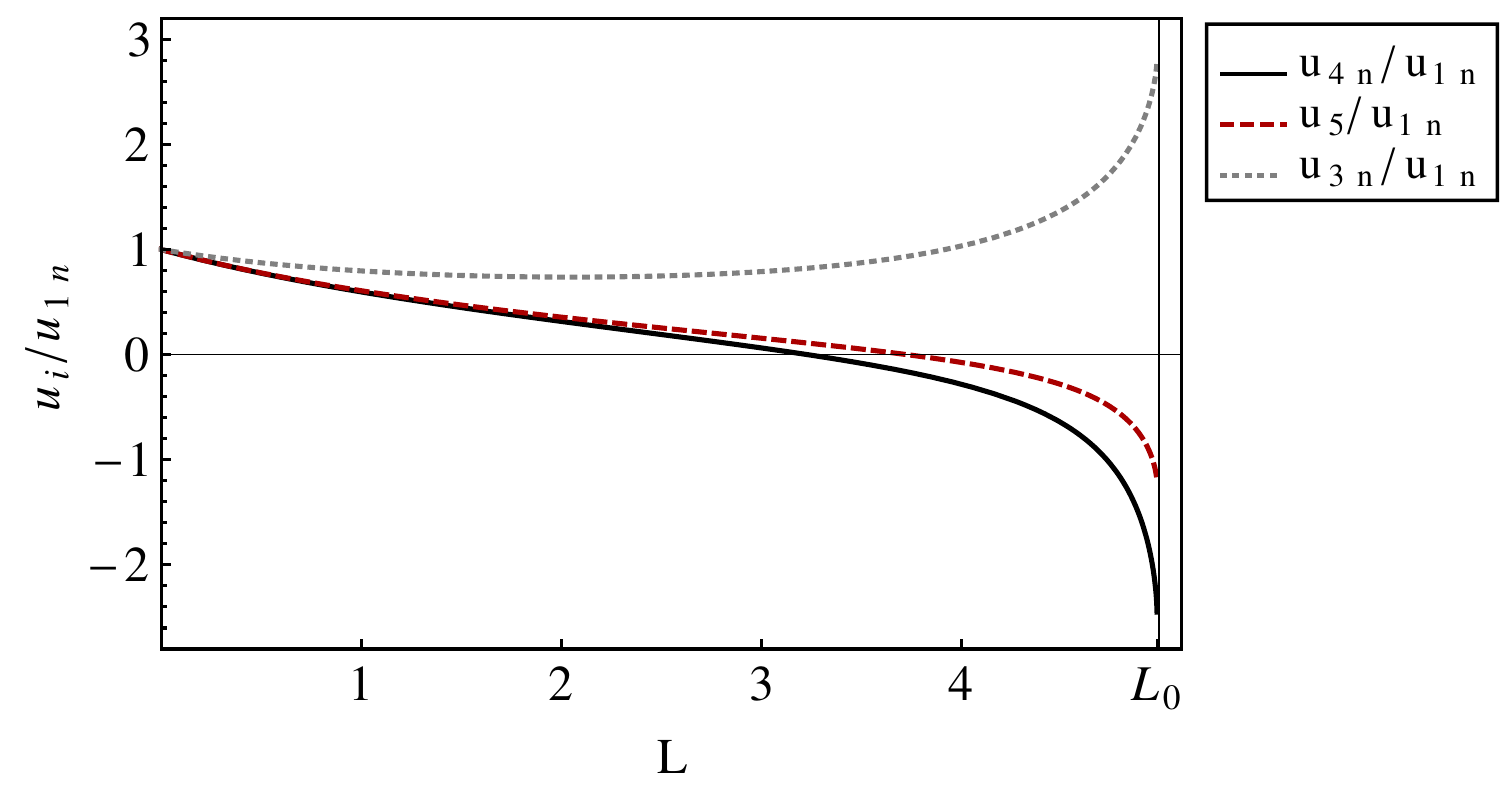}
 \caption{Ratios of couplings for the flow to the 3-pocket fixed trajectory (3p) in the toy model for bare values $U=U'=0.1/N_F$, $J=J'=0.03/N_F$ and $a=a_n=1$, $N_F$ is the density of states on the FSs. All ratios tend to zero (upper penal), except for those within the triad of electron pockets and the $M-$centered hole pocket.}
\label{fig:ratios}
\end{figure}
We searched for different fixed trajectories of  Eq.~(\ref{eq:rgxy}) along which the couplings diverge, but their ratios tend to fixed values. This can be seen in Fig.~\ref{fig:ratios}. Accordingly, we single out one of the coupling, say $u_0$, and write all other couplings as
\begin{equation}
u_i=\gamma_i u_0,
\end{equation}
Along the fixed trajectory, $u_0$ flows to infinity, but $\gamma_i$ tend to finite values.
 Solving for the fixed trajectory of the set of  coupled pRG equations,  Eq.~(\ref{eq:rgxy}), then reduces to finding the fixed point solution of
\begin{equation}
\beta_i:=\partial_L\gamma_i=\frac{1}{u_i}(\partial_Lu_i-\gamma_i\partial_Lu_0)=0.
\end{equation}
The fixed trajectory is stable if small perturbations around the fixed point do not grow, i.e.~the stability matrix $\partial \beta_i/\partial \gamma_j|_{\gamma^*}$, which describes the linearized flow around the fixed point, should have only negative eigenvalues. For the toy model we find two stable fixed trajectories, separated by a fixed point solution with a single unstable direction.  In the main text we labeled the two stable fixed trajectories  as effective 4-pocket model (4p) and effective 3-pocket mode (3p). The behavior of the couplings along these two stable fixed trajectories  is
\begin{enumerate}[(1)]
 \item 4p
\begin{align}
u_i&=\gamma_i u_1\nonumber\\
u_1&=\frac{1}{1+\gamma_3^2/a^2}\frac{1}{L_0-L}\nonumber\\
\gamma_2&=\gamma_{1n}=\gamma_{2n}=\gamma_{3n}=\gamma_{4n}=\gamma_c=0\nonumber\\
\gamma_3&=\pm a\sqrt{8a^2-1+4\sqrt{1-a^2+4a^4}}\nonumber\\
\gamma_4&=\gamma_5=1-2a^2-\sqrt{1-a^2+4a^4}
\end{align}
\item 3p
\begin{align}
u_i&=\gamma_i u_{1n}\nonumber\\
u_{1n}&=\frac{1}{1+\gamma_{3n}^2/a_n^2}\frac{1}{L_0-L}\nonumber\\
\gamma_{2n}&=\gamma_c=\gamma_{1}=\gamma_{2}=\gamma_{3}=\gamma_{4}=0\nonumber\\
\gamma_{3n}&=\pm a_n\sqrt{4a_n^2-1+2\sqrt{4-2a_n^2+4a_n^4}}\nonumber\\
\gamma_{4n}&=2\gamma_5=2-2a_n^2-\sqrt{4-2a_n^2+4a_n^4}
\end{align}
\end{enumerate}
Because the bare values for $\gamma_3$, $\gamma_{3n}$ are positive, the system reaches the stable FT  with positive $\gamma_3$, $\gamma_{3n}$.
  We see that along the stable fixed trajectories,  either all $\gamma_i$ for interactions with the $\Gamma-$centered hole pockets vanish (3p),
  or all $\gamma_i$ for the interactions with the third hole pocket at $M$ vanish (4p). This does not mean that the interactions themselves vanish, it only means that these interactions do not grow as fast as other interactions.  These couplings actually still increase under pRG but with  exponents smaller than one.
  This means that, to leading order, the system flows to either 4-pocket model (4p)  or 3-pocket model (3p).   However the subleading terms still have an impact on the emergent order, as they determine how the order parameter behaves at the remaining hole pocket(s).

   The third, weakly unstable fixed trajectory is symmetry-enhanced in the sense that $u_1=u_{1n}$, and $u_3/a=u_{3n}/a_n$. Along this trajectory
    all ratios (except for $\gamma_2,\gamma_{2n}$) attain finite values.  Specifically, we obtain
\begin{align}
u_i&=\gamma_i u_1\nonumber\\
u_1&=\frac{1}{1+\gamma_3^2/a^2}\frac{1}{L_0-L}\nonumber\\
\gamma_{1n}&=\gamma_1=1 \quad \frac{\gamma_{3n}}{a_n}=\pm\frac{\gamma_3}{a} \quad \gamma_2=\gamma_{2n}=0\nonumber\\
\gamma_3&=\sqrt{8a^4-a^2+4a^2a_n^2+a^2\sqrt{15+(8a^2+4a_n^2-1)^2}}\nonumber\\
\gamma_5&=1-a_n^2-2a^2-\sqrt{1+a_n^4-a^2+4a^4+a_n^2(4a^2-\frac{1}{2})}\nonumber\\
\gamma_c&=\pm a_n\frac{2a-\sqrt{2\gamma_3^2a_n^2+4a^2(1+\gamma_3^2)}}{a_n^2+2a^2}\nonumber\\
\gamma_4&=\pm\frac{a_n}{2a}\gamma_c+\frac{\gamma_3^2}{4a^2}-\frac{3}{4}\nonumber\\
\gamma_{4n}&=\mp\frac{2a}{a_n}\gamma_c+\gamma_5
\label{2}
\end{align}
For $a=a_n=1$ $\gamma_i$ in (\ref{2}) reduce to
\begin{align}
\gamma_3&=\pm\gamma_{3n} \quad \gamma_4=\gamma_{4n}=\pm\gamma_c\nonumber\\
\gamma_3&=\sqrt{11+2\sqrt{34}} \quad \gamma_4=-\frac{1}{3}(4+\sqrt{34}) \nonumber\\
\gamma_5&=-2-\sqrt{\frac{17}{2}}
\end{align}

\begin{figure}[t!]
\centering
 \includegraphics[width=.7\columnwidth]{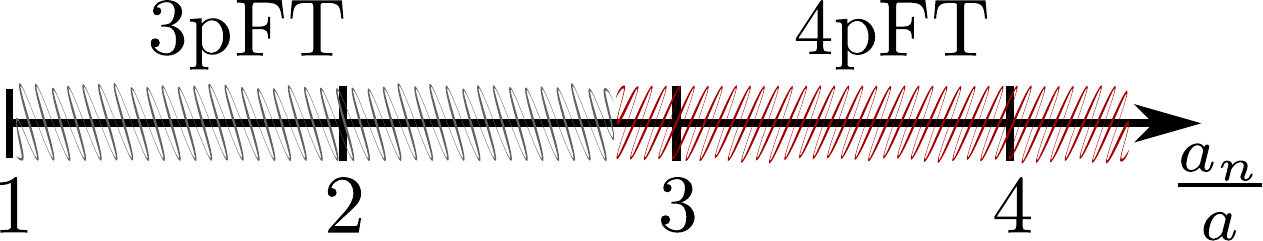}
 \caption{Fixed trajectory at the end of the flow for different values of $a_n/a$. 3pFT (4pFT) denotes the effective 3p (4p) model. Bare values are $U=\pr{U},J=\pr{J}$ and $J/U=0.3$.}
\label{fig:C3simpl}
\end{figure}

Like we said, this fixed trajectory has one unstable direction when we consider deviations from it. We verified that, depending on the sign of deviation along the unstable direction, the system flows  either to one or to the other stable fixed trajectory.
We present the phase diagram for different bare values  in the main text and here present the result of our study of the stability regimes of 4p and 3p  at various
 $a_n/a$ in Fig.\ref{fig:C3simpl}.

%-------------------------------------------------------------------------------------------------

%--------------------------------------------------------------------------------------------------------------------------------------------------------------
\subsubsection{Susceptibilities}

To decide which order wins  and develops at low energies, we introduce vertices $\Gamma_i$ that describe the coupling between fermions and order parameters.
The vertices in turn determine the susceptibilities in the corresponding ordering channel, whose divergence would signal a phase transition.
Here we focus on SDW, CDW, and SC channels.  The analysis of the susceptibilities in the Pomeranchuk channels is discussed afterwards.

\begin{figure*}[t]
\centering
 \includegraphics[width=1.9\columnwidth]{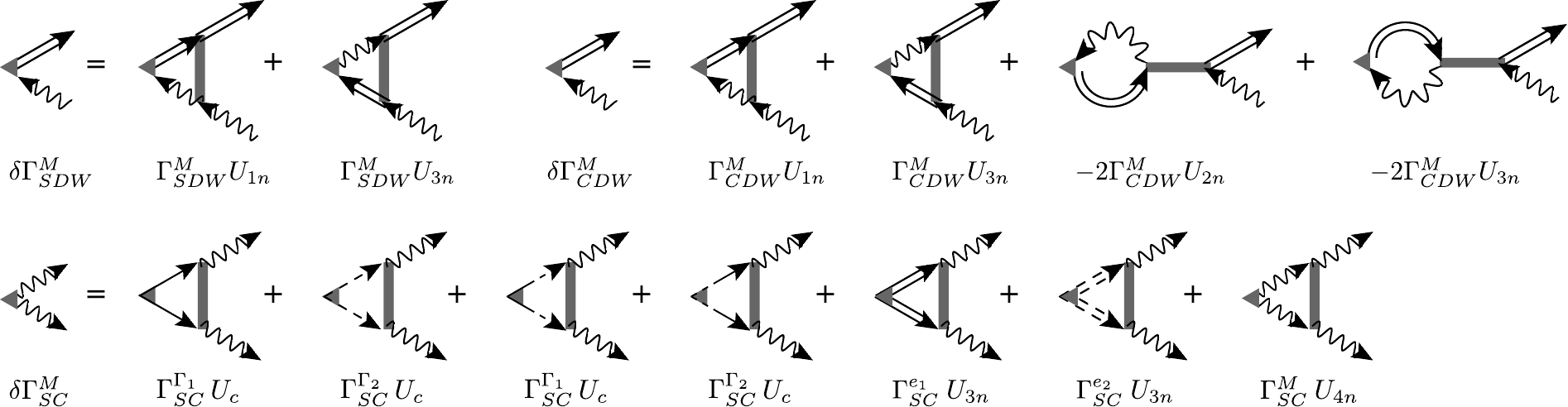}
 \caption{Diagrammatic representation of the 1-loop renormalization of representative SDW, CDW, and SC vertices. In the RG equations of the superconducting vertices, only the combinations $\Gamma_{SC}^\Gamma:=\Gamma_{SC}^{\Gamma_1}+\Gamma_{SC}^{\Gamma_2}, \Gamma_{SC}^e:=\Gamma_{SC}^{e_1}+\Gamma_{SC}^{e_2}$ appear.}
\label{fig:loopsGammas}
\end{figure*}

The vertices are renormalized by the corresponding polarization bubbles and diverge with a certain exponent when the running couplings approach the fixed trajectory, as  $\Gamma_i\propto (L_0-L)^{-\beta}$. The  susceptibilities
\begin{align}
\chi_i- \chi_0\propto \int_L~d\pr{L}\Gamma_i^2(\pr{L}),
\end{align}
 then behave as $\chi\propto(L_0-L)^{1-2\beta} + \text{const}$.  In order to diverge the vertex exponent must satisfy $\beta\geq1/2$.
 The one-loop renormalization of the vertices are shown in Fig.~\ref{fig:loopsGammas}. In analytic form, the pRG equations for the vertices in the SDW and CDW channels are
 \begin{equation}
\begin{aligned}
\partial_L\Gamma_{SDW}^{\Gamma}&=\rund{u_1+\frac{u_3}{a}}\Gamma_{SDW}^{\Gamma}\\
\partial_L\Gamma_{SDW}^{M}&=\rund{u_{1n}+\frac{u_{3n}}{a_n}}\Gamma_{SDW}^{M}\\
\partial_L\Gamma_{CDW}^{\Gamma}&=\rund{u_1-2u_2-\frac{u_3}{a}}\Gamma_{CDW}^{\Gamma}\\
\partial_L\Gamma_{CDW}^{M}&=\rund{u_{1n}-2u_{2n}-\frac{u_{3n}}{a_n}}\Gamma_{CDW}^{M}.
\end{aligned}
\end{equation}
By inserting the values for the fixed trajectories, we obtain the exponents $\beta_i$:
 \begin{equation}
\begin{aligned}
\beta_{SDW}^{(4p)}&=\frac{1+\gamma_3/a}{1+\gamma_3^2/a^2} \quad \beta_{CDW}^{(4p)}=\frac{1-\gamma_3/a}{1+\gamma_3^2/a^2}\\
\beta_{SDW}^{(3p)}&=\frac{1+\gamma_{3n}/a_n}{1+\gamma_{3n}^2/a_n^2} \quad \beta_{CDW}^{(3p)}=\frac{1-\gamma_{3n}/a_n}{1+\gamma_{3n}^2/a_n^2}.
\end{aligned}
\end{equation}
Note that $\gamma_3,\gamma_{3n}$ also depend on $a,a_n$ in these expressions. The exponents attain their maximal values at $a=1,a_n=1$ with $\beta_{SDW}^{(4p)}\approx0.30,~\beta_{CDW}^{(4p)}\approx-0.18$ and $\beta_{SDW}^{(3p)}\approx0.43,~\beta_{CDW}^{(3p)}\approx-0.20$. These values do not lead to a divergent susceptibility, i.e. the corresponding order does not develop if the normal state becomes unstable before the Fermi energy is reached.

The pRG flow of the vertices in the particle-particle channel obeys
\begin{align}\label{eq:SC}
\partial_L \begin{pmatrix} \Gamma_{SC}^{e} \\ \Gamma_{SC}^{\Gamma} \\ \Gamma_{SC}^{M}\end{pmatrix} =\begin{pmatrix}-2u_5 & -2u_3 & -2u_{3n} \\ -2u_3 & -2u_4 & -2u_c \\ -u_{3n} & -u_c& -u_{4n} \end{pmatrix}\begin{pmatrix}\Gamma_{SC}^{e}\\ \Gamma_{SC}^{\Gamma}  \\ \Gamma_{SC}^{M}\end{pmatrix},
\end{align}
where we have absorbed different prefactors into $\Gamma_{SC}$ as  $\sqrt{A_h/A_e}\Gamma_{SC}^\Gamma\rightarrow\Gamma_{SC}^\Gamma,\sqrt{A_M/A_e}\Gamma_{SC}^M\rightarrow\Gamma_{SC}^M$. For 4pFT and 3pFT, this set reduces to a 2x2 matrix,
 and the diagonalization of Eq.~(\ref{eq:SC}) yields in these two cases
 \begin{equation}
\begin{aligned}
\beta_{SC,+-/++}^{(4p)}&=\frac{-\gamma_4-\gamma_5\pm\sqrt{(\gamma_4-\gamma_5)^2+4\gamma_3^2/a^2}}{1+\gamma_3^2/a^2}\\
\beta_{SC,+-/++}^{(3p)}&=\frac{-\gamma_{4n}-2\gamma_5\pm\sqrt{(\gamma_{4n}-2\gamma_5)^2+8\gamma_{3n}^2/a_n^2}}{1+\gamma_{3n}^2/a_n^2}.
\end{aligned}
\end{equation}
The largest eigenvalues correspond to the $s_{+-}$ superconducting state and satisfy $\beta_{SC,+-}>1/2$. For $a=a_n=1$ they are $\beta_{SC,+-}^{(4p)}= 0.86$, and $\beta_{SC,+-}^{(3p)} = 0.72$.  Because these $\beta_{SC}$ are larger than $1/2$, we find that the system develops superconductivity at low energies rather than SDW or CDW order.
 From the analysis of the fixed trajectory we can infer that the gap changes sign either between  the electron pockets and the two $\Gamma-$centered hole pockets (for 4p), or between the  electron pockets and the $M-$centered hole pocket (for 3p). In both cases,  this is conventional $s^{+-}$ gap structure.

   To determine the sign of the superconducting gap on the remaining hole pocket(s), we must include the residual interactions (the once which diverge with smaller exponents). To do this and to verify our analytical reasoning, we solved the set of pRG equations for the couplings and the set of the vertices in the SC channel, Eq.~(\ref{eq:SC}), numerically.
      We find two positive (attractive)  and one negative eigenvalue in the SC channel.  The negative one obviously corresponds to repulsive interaction in $s^{++}$ channel.
        The positive eigenvalues correspond to $s^{+-}$ gap structure. For the largest positive eigenvalue along the 3p FT or 4p FT
         the gap(s) on the remaining hole pocket(s) align such that the sign of the gap on all three hole pockets is the same (and opposite to the gap sign on the two electron pockets).  This is the "conventional'' $s^{+-}$ gap structure.  However, the size of the vertex, which is related to the gap size, on the residual pocket is smaller than on the dominant pockets.
         The smaller positive eigenvalue along the 3p FT or the 4p FT actually starts negative at small $L$ and then changes the sign in the process of the RG flow.
          For the 4p FT,  the gap structure that corresponds to this eigenvalue
            has the same sign of the gap on the $M-$centered hole pocket as on the electron pockets, i.e., there is one sign of the gap on the two $\Gamma-$centered hole pockets and another sign on the other three pockets.
              For the 3p FT and for this eigenvalue, the sign of the gap on the $\Gamma-$centered hole pockets and on the electron pockets is the same, and opposite to that on the $M-$ hole pocket.  The gap structure of this kind was proposed in Ref. \cite{comm_kot} and termed as "orbital anti-phase''.
               Our RG analysis shows that along the fixed trajectory such a state is subleadng to a conventional $s^{+-}$.
                Finally, we computed the gap structure along the  weakly unstable FT  of Eq. (\ref{2}) and found that it is also a conventional $s^{+-}$. We do not find d-wave order.

\begin{figure*}[t]
\centering
 \includegraphics[width=1.9\columnwidth]{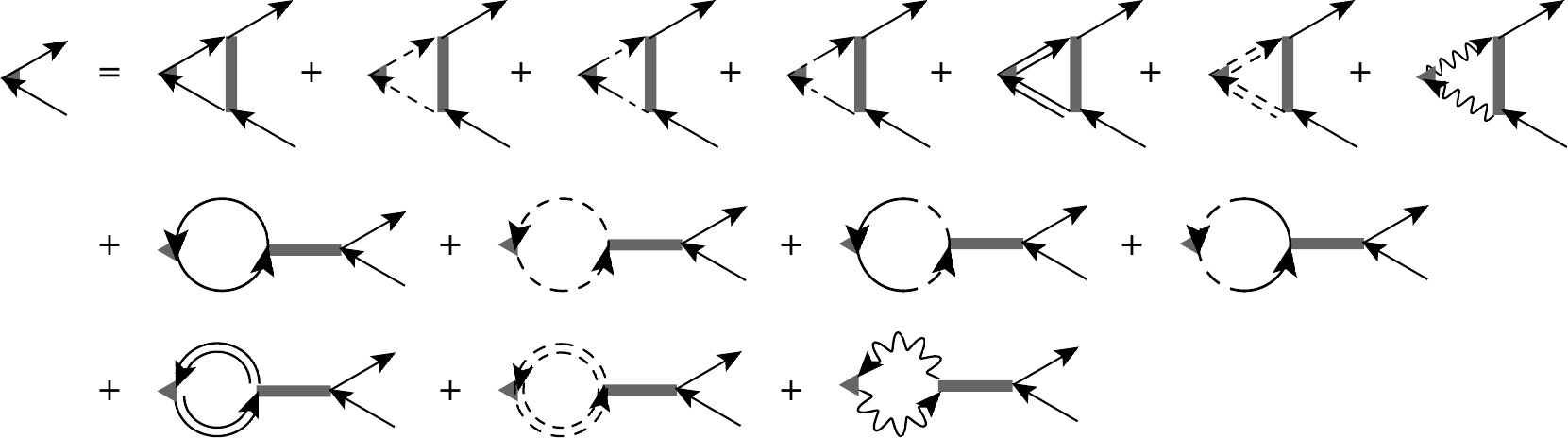}
 \caption{Diagrammatic representation of the 1-loop renormalization of a representative Pomeranchuk vertex corresponding to the orbital density $n_{xz}$. The polarization bubbles are not logarithmic as they involve two identical propagators.}
\label{fig:Pomvertex}
\end{figure*}

To analyze orbital ordering, we calculate the vertices and susceptibilities in the Pomeranchuk channel with the orbital densities $n_\mu = d_{\mu}^\dagger d_\mu$ as order parameters. The analysis is somewhat different than before because the polarization bubbles that renormalize the Pomeranchuk vertices are not logarithmically divergent as can be seen in Fig.~\ref{fig:Pomvertex} . However, the scale-dependence of the interaction provides a logarithmic renormalization. Summing only logarithmic terms then leads to pRG equations in the Pomeranchuk channel of the form $\partial_L \Gamma_\mu\propto \Gamma_{\mu}^0\partial_L u$, i.e. $\Gamma_\mu\propto\Gamma^0_\mu(1+u)$. Since the couplings flow as $u\propto(L_0-L)^{-1}$, the Pomeranchuk vertex grows with exponent $\beta_\mu=1$ and overtakes the SC vertex at the end of the flow. Note however that the renormalization of the Pomeranchuk vertex develops when the couplings become of order one so that corrections to 1-loop RG may contribute. Explicitly the pRG equation of the Pomeranchuk channel for the toy model reads
\begin{widetext}
\begin{align}\label{eq:Pom}
\arraycolsep=1.4pt
\frac{d}{dL}  \!\begin{pmatrix} \Gamma_{xz}^{\Gamma} \\ \Gamma_{yz}^{\Gamma} \\ \Gamma_{xy}^{X} \\ \Gamma_{xy}^{Y} \\ \Gamma_{xy}^{M}\end{pmatrix} \!=\!
-2\frac{d}{dL}\!\begin{pmatrix} u_4 & 0 & \frac{A_e}{A}(2u_1-u_2) &  \frac{A_e}{A}(2 u_1- u_2) &  0\\ 0 & u_4 &  \frac{A_e}{A}(2u_1-u_2) &  \frac{A_e}{A}(2u_1-u_2) & 0 \\  \frac{A_h}{A}(2u_1-u_2) & \frac{A_h}{A}(2u_1-u_2) & u_5& 0 & \frac{A_M}{A_n}(2u_{1n}-u_{2n}) \\ \frac{A_h}{A}(2u_1-u_2) & \frac{A_h}{A}(2u_1-u_2) & 0& u_5 & \frac{A_M}{A_n}(2u_{1n}-u_{2n}) \\ 0 & 0 & \frac{A_e}{A_n} (2 u_{1n}- u_{2n}) & \frac{A_e}{A_n} (2 u_{1n}- u_{2n}) & u_{4n}\end{pmatrix}
\begin{pmatrix}\Gamma_{xz}^{\Gamma} \\ \Gamma_{yz}^{\Gamma} \\ \Gamma_{xy}^{X} \\ \Gamma_{xy}^{Y} \\ \Gamma_{xy}^{M}\end{pmatrix},
\end{align}
\end{widetext}
where we have omitted the irrelevant couplings (Eq.~(\ref{eq:irrelevant})) and set $m_c=m_d$. As has been already obtained in Ref.~\cite{we_last}, the leading instability in the Pomeranchuk channel along the 4pFT is in the d-wave channel with non-equal densities $n_{xz}-n_{yz}$. Along the 3pFT an instability with different densities on the electron pockets $n_{xy}(X)-n_{xy}(Y)\neq0$ develops, which also breaks $C_4$ symmetry. Such an order splits one of the band degeneracies of the electron bands in the folded Brillouin zone.

Finally, we comment on the system behavior in a situation when the system does not reach a fixed trajectory before the RG scale $L$ becomes comparable to $L_F = \log{\Lambda/E_F}$.  Because the susceptibility in the SDW channel is the largest  over a wide range of $L$, it is most likely that in this situation the system develops an SDW order.  We compared the behavior of SDW vertices involving fermions from one of the electron pockets and either fermions from $\Gamma-$centered hole pockets ($\Gamma^\Gamma_{SDW}$) or from the $M-$pocket ($\Gamma^M_{SDW}$).  We found that  $\Gamma_{SDW}^M>\Gamma_{SDW}^\Gamma$ if the flow is towards the 3pFT, and
$\Gamma_{SDW}^\Gamma>\Gamma_{SDW}^M$  if the  flow is towards  the 4pFT.  This implies that in the first case SDW order predominantly involves the triad of
two electron pockets and the $M$ hole pockets, while in the second case it involves two electron pockets and two $\Gamma$-centered hole pockets.

%-------------------------------------------------------------------------------------------------
\subsection{PRG for full 5-pocket model}\label{sec:fullprg}
\subsubsection{PRG equations and fixed trajectories}
We now move to the full 5-band model with  $xz/yz$ orbital content on the electron pockets. Like we said, in this case we have
 19 more couplings (the total number of the couplings is 40).
 The couplings $\tilde U_4,\tilde{\tilde U}_4,\tilde U_5,\tilde{\tilde U}_5,U_a,U_b$ do not couple to additional terms and continue to flow to zero under pRG.
  We find six additional couplings $\tilde V_5$, $\tilde{\tilde V}_5$, $V_a,  V_b, \bar V_a, \bar V_b$
  that flow to zero. The corresponding  pRG equations are
\begin{align}
4\pi \frac{d}{dL}\rund{\tilde V_5 \pm \tilde{\tilde V}_5}&=-c^{(4)}_{pp}\rund{\tilde V_5 \pm \tilde{\tilde V}_5}^2\notag\\
4\pi \frac{d}{dL}\rund{ V_a \pm V_b}&=-c^{(5)\pm}_{pp}\rund{V_a \pm V_b}^2\notag\\
4\pi \frac{d}{dL}\rund{ \bar V_a \pm \bar V_b}&=-c^{(6)}_{pp}\rund{\bar V_a \pm \bar V_b}^2, \label{eq:irrelevant2}
\end{align}
where $c^{(4)}_{pp}=1/L \int d\omega\int d^2k G_{f_{31},f_{31}}G_{f_{32},f_{32}}$, $c^{(5)\pm}_{pp}=1/L \int d\omega\int d^2k (G_{f_{31},f_{31}}G_{f_{1},f_1}\pm G_{f_{31},f_{1}}G_{f_{1},f_{31}})$, and $c^{(6)}_{pp}=1/L \int d\omega\int d^2k G_{f_{31},f_{31}}G_{f_{2},f_2}$. For the other couplings  we make the same conjecture as for the toy model, i.e.,
   assume that for stable and weakly unstable fixed trajectories  $U_i=\bar U_i,V_i=\bar V_i$.
     The one-loop RG equations for the remaining dimensionless couplings are
\begin{align}\label{eq:rgfull}
\dot u_1&=u_1^2+\frac{u_3^2}{a^2}\nonumber\\
\dot u_{1n}&=u_{1n}^2+\frac{u_{3n}^2}{a_n^2}\nonumber\\
\dot u_2&=2u_2(u_1-u_2)\nonumber\\
\dot u_{2n}&=2u_{2n}(u_{1n}-u_{2n})\nonumber\\
\dot u_3&=2u_3(2u_1-u_2-u_5)-2bu_3u_{4}-u_{3n}u_c \nonumber\\
&- 2v_3v_c - 2H (u_3v_c + v_3 u_5)\nonumber\\
\dot u_{3n}&=2u_{3n}(2u_{1n}-u_{2n}-u_5)-u_{3n}u_{4n}-2bu_3 u_c \nonumber\\
&- 2v_{3n}v_c - 2H(u_{3n}v_c+v_{3n}u_5)\nonumber\\
\dot u_4&=-2bu_4^2-2u_3^2-2u_c^2 - 2v_3^2-4H u_3 v_3\nonumber\\
\dot u_{4n}&=-u_{4n}^2-2u_{3n}^2-2bu_c^2-2v_{3n}^2-4Hu_{3n}v_{3n}\nonumber\\
\dot u_5&=-2u_5^2-2bu_3^2-u_{3n}^2-2v_c^2-4Hu_5v_c\nonumber\\
\dot u_c&=-2bu_4u_c-u_{4n}u_c-2u_3u_{3n}\nonumber\\
&-2v_3v_{3n}-2H(v_3u_{3n}+v_{3n}u_3)\nonumber\\
\dot v_1&=v_1^2+\frac{v_3^2}{c^2}\nonumber\\
\dot v_{1n}&=v_{1n}^2+\frac{v_{3n}^2}{c_n^2}\nonumber\\
\dot v_2&=2v_2(v_1-v_2)\nonumber\\
\dot v_{2n}&=2v_{2n}(v_{1n}-v_{2n})\nonumber\\
\dot v_3&=2v_3(2v_1-v_2-v_5)-2bv_3u_{4}-v_{3n}u_c \nonumber\\
&- 2u_3v_c - 2H (v_3v_c + u_3 v_5)\nonumber\\
\dot v_{3n}&=2v_{3n}(2v_{1n}-v_{2n}-v_5)-v_{3n}u_{4n}-2bv_3 u_c \nonumber\\
&- 2u_{3n}v_c - 2H(v_{3n}v_c+u_{3n}v_5)\nonumber\\
\dot v_5&=-2v_5^2-2bv_3^2-v_{3n}^2-2v_c^2-4Hv_5v_c\nonumber\\
\dot v_c&=-2bv_3u_3-2v_{5}v_c-2u_5v_{c}\nonumber\\
&-2v_{3n}u_{3n}-2H(v_c^2+v_{5}u_5),
\end{align}
where the additional parameters are
\begin{align}
c&=\frac{\sqrt{A_h A_{e}'}}{A'} \quad c_n=\frac{\sqrt{A_M A_{e}'}}{A_n'} \quad H=\frac{A_e^a}{\sqrt{A_eA_e'}}\nonumber\\
A_e&=\frac{1}{L}\frac{1}{(2\pi)^2}\int d\omega\int d^2k G_{f_{1},f_{1}}G_{f_{1},f_{1}}%\nonumber\\
=m_e\int \frac{d\theta}{2\pi} \cos^4\varphi_1\nonumber\\
A_e'&=\frac{1}{L}\frac{1}{(2\pi)^2}\int d\omega\int d^2k G_{f_{31},f_{31}}G_{f_{31},f_{31}}\nonumber\\
&=m_e\int \frac{d\theta}{2\pi} \sin^4\varphi_1\nonumber\\
A'&=\frac{1}{L}\frac{1}{(2\pi)^2}\int d\omega\int d^2k G_{f_{31},f_{31}}G_{d_1,d_1}\nonumber\\
&=2\int \frac{d\theta}{2\pi}\sin^2\varphi_1\rund{\frac{m_c m_e}{m_c+m_e}\cos^2\theta+\frac{m_d m_e}{m_d+m_e}\sin^2\theta}\nonumber\\
A_n'&=\frac{1}{L}\frac{1}{(2\pi)^2}\int d\omega\int d^2k G_{f_{31},f_{31}}G_{M}\nonumber\\
&=2\frac{m_Mm_e}{m_M+m_e}\int\frac{d\theta}{2\pi} \sin^2\varphi_1\nonumber\\
A_e^a &=\frac{1}{L}\frac{1}{(2\pi)^2}\int d\omega\int d^2k G_{f_1,f_{31}}G_{f_{1},f_{31}}\nonumber\\
&=m_e\int\frac{d\theta}{2\pi}\sin^2\varphi_1\cos^2\varphi_1.
\end{align}

Interestingly, we find that the stable fixed trajectories of the full model  lead to  the same decoupling  at low-energies
into effective three or four pocket models, as in the toy model. In distinction to the toy model, however, now  there are two 3p and two 4p effective models ($3_1, 3p_2, 4p_1, 4p_2$).  These four stable fixed trajectories are specified by
 \begin{enumerate}[(4p$_1$)]
 \item[(4p$_1$)]
\begin{align}
u_i&=\gamma_i u_1 \quad v_i=g_i u_1\nonumber\\
u_1&=\frac{1}{1+\gamma_3^2/a^2}\frac{1}{L_0-L}\nonumber\\
\gamma_3&=\pm a\sqrt{8a^2-1+4\sqrt{1-a^2+4a^4}}\nonumber\\
\gamma_4&=\gamma_5=1-2a^2-\sqrt{1-a^2+4a^4}
\label{3}
\end{align}
 \item[(4p$_2$)]
\begin{align}
u_i&=\gamma_i v_1 \quad v_i=g_i v_1\nonumber\\
v_1&=\frac{1}{1+g_3^2/c^2}\frac{1}{L_0-L}\nonumber\\
g_3&=\pm c\sqrt{8c^2-1+4\sqrt{1-c^2+4c^4}}\nonumber\\
\gamma_4&=g_5=1-2c^2-\sqrt{1-c^2+4c^4}
\label{4}
\end{align}
 \item[(3p$_1$)]
\begin{align}
u_i&=\gamma_i u_{1n} \quad v_i=g_i u_{1n}\nonumber\\
u_{1n}&=\frac{1}{1+\gamma_{3n}^2/a_n^2}\frac{1}{L_0-L}\nonumber\\
\gamma_{3n}&=\pm a_n\sqrt{4a_n^2-1+2\sqrt{4-2a_n^2+4a_n^4}}\nonumber\\
\gamma_{4n}&=2\gamma_5=2-2a_n^2-\sqrt{4-2a_n^2+4a_n^4}
\label{5}
\end{align}
 \item[(3p$_2$)]
\begin{align}
u_i&=\gamma_i v_{1n} \quad v_i=g_i v_{1n}\nonumber\\
v_{1n}&=\frac{1}{1+g_{3n}^2/c_n^2}\frac{1}{L_0-L}\nonumber\\
g_{3n}&=\pm c_n\sqrt{4c_n^2-1+2\sqrt{4-2c_n^2+4c_n^4}}\nonumber\\
g_{4n}&=2g_5=2-2c_n^2-\sqrt{4-2c_n^2+4c_n^4}
\label{6}
\end{align}
\end{enumerate}
All couplings not presented in the above formulas evolve with smaller exponents.
 Note that the ratios of the couplings in Eqs. (\ref{3}-\ref{6}) do not depend on the parameter $H$.

 We see from  Eqs. (\ref{3}-\ref{4}) that for 4p$_{1}$ and 4p$_{2}$ all interactions involving the $M$-centered hole pocket become subleading, like in the toy model.
    For 4pFT$_{1}$ the interactions involving $xz/yz$  orbital components on the electron pockets become leading compared to the interactions involving $xy$ orbital components, i.e., to first approximation the two electron pockets can be approximated as $xz/yz$-pockets. For 4p$_{2}$ the situation is opposite -- the interactions involving $xy$  orbital component on the electron pockets become dominant compared to the interactions involving $xz/yz$ orbital components, i.e., to first approximation the two electron pockets can be approximated as $xy$ pockets.  These two fixed trajectories have been analyzed in Ref. \cite{we_last}.
     The situation is equivalent for the 3p$_{1}$ and 3p$_{2}$, see Eqs. (\ref{5}-\ref{6}).
     In the first case,  the interactions involving $xz/yz$  orbital component on the electron pockets become leading, and in the second  the interactions involving $xy$  orbital component on the electron pockets become leading.

     These different effective low-energy models  are sketched in Fig. 1 in the main text. We also note that the behavior of different couplings along
       4p$_{1}$ and 4p$_{2}$ are quite similar, see Eqs. (\ref{3}, \ref{4}), and the same is true for the couplings along  3p$_{1}$ and 3p$_{2}$,  Eqs. (\ref{5},\ref{6}).  Whether the system flows to
        4p$_{1}$ or 4p$_{2}$ (or to  3p$_{1}$ or 3p$_{2}$) depends on the initial values of the couplings.

The stable FTs are separated by several weakly unstable ones with only a single direction along which perturbations grow.
  For general $a,a_n,c,c_n$, and $H$ we determined these FTs and checked their stability numerically. For $a=a_n=c=c_n=1$ these weakly unstable FTs can be analyzed
    analytically. The FTs with only one unstable direction are (the notations are self-evident):
\begin{enumerate}[(4p$_1$+4p$_2$)]
 \item[(4p$_1$+4p$_2$)]
 \begin{equation}
\begin{aligned}
u_i&=\gamma_i u_1 \quad v_i=g_i u_1\\
u_1&=v_1 \quad u_3=v_3 \quad u_5=v_5=v_c\\
u_1&=\frac{1}{1+\gamma_3^2/a^2}\frac{1}{L_0-L}\\
\gamma_3&=\pm\sqrt{15+16H+4\sqrt{15+30H+16H^2}}\\
\gamma_4&=2(H+1)\gamma_5=-3-4H\sqrt{15+30H+16H^2}
\end{aligned}
\end{equation}
\item[(3p$_1$+3p$_2$)]
\begin{equation}
\begin{aligned}
u_i&=\gamma_i u_{1n} \quad v_i=g_i u_{1n}\\
u_{1n}&=v_{1n} \quad u_{3n}=v_{3n} \quad u_5=v_5=v_c\\
u_{1n}&=\frac{1}{1+\gamma_{3n}^2/a_n^2}\frac{1}{L_0-L}\\
\gamma_{3n}&=\pm \sqrt{7+8H+4\sqrt{4+7H+4H^2}}\\
\gamma_{4n}&=4(H+1)\gamma_5=-2-4H-2\sqrt{4+7H+4H^2}
\end{aligned}
\end{equation}
\item[(3p$_1$+4p$_1$)]
 \begin{equation}
\begin{aligned}
u_i&=\gamma_i u_1 \quad v_i=g_i u_1\\
u_1&=u_{1n} \quad u_3=\pm u_{3n} \quad u_4=u_{4n}=\pm u_c\\
u_1&=\frac{1}{1+\gamma_3^2/a^2}\frac{1}{L_0-L}\\
\gamma_3&=\sqrt{11+2\sqrt{34}} \quad \gamma_4=-\frac{1}{3}(4+\sqrt{34}) \\
\gamma_5&=-2-\sqrt{\frac{17}{2}}
\end{aligned}
\end{equation}
\item[(3p$_2$+4p$_2$)]
 \begin{equation}
\begin{aligned}
u_i&=\gamma_i v_{1} \quad v_i=g_i v_{1}\\
v_1&=v_{1n} \quad v_3=\pm v_{3n} \quad u_4=u_{4n}=\pm u_c\\
g_3&=\sqrt{11+2\sqrt{34}} \quad \gamma_4=-\frac{1}{3}(4+\sqrt{34}) \\
g_5&=-2-\sqrt{\frac{17}{2}}.
\end{aligned}
\end{equation}
\end{enumerate}
Again all couplings not listed in the formulas above have smaller exponents.  A detailed analysis of the structure of weakly unstable FTs in the full 4-pocket model is presented in Ref.~\cite{we_future}.

 Finally there is a high-symmetry FT with two unstable directions. Along this FT  all couplings are non-zero:
 \begin{equation}
\begin{aligned}
u_i&=\gamma_i u_1 \quad v_i=g_i u_1\\
u_1&=u_{1n}=v_1=v_{1n} \quad u_3=u_{3n}=v_3=v_{3n}\\
u_4&=h_{4n}=u_c \quad u_5=v_5=v_c\\
u_1&=\frac{1}{1+\gamma_3^2/a^2}\frac{1}{L_0-L}\\
\gamma_3&=\pm\sqrt{23+24H+4\sqrt{34+69H+36H^2}}\\
\gamma_4&=\frac{4}{3}(H+1)\gamma_5=-\frac{10}{3}-4H-\frac{2}{3}\sqrt{34+69H+36H^2}.
\end{aligned}
\end{equation}

\subsubsection{Susceptibilities}

As in the toy model, we introduce vertices that couple to different order parameter fields to determine  which order develops first at low energies.
 In the SDW channel, we now have four vertices
 \begin{equation}
\begin{aligned}
\partial_L\Gamma_{SDW}^{\Gamma,1}&=\rund{u_1+\frac{u_3}{a}}\Gamma_{SDW}^{\Gamma,1}\\
\partial_L\Gamma_{SDW}^{\Gamma,2}&=\rund{v_1+\frac{v_3}{c}}\Gamma_{SDW}^{\Gamma,2}\\
\partial_L\Gamma_{SDW}^{M,1}&=\rund{u_{1n}+\frac{u_{3n}}{a_n}}\Gamma_{SDW}^{M,1}\\
\partial_L\Gamma_{SDW}^{M,2}&=\rund{v_{1n}+\frac{v_{3n}}{c_n}}\Gamma_{SDW}^{M,2}\\
\end{aligned}
\end{equation}
 where
  indices $1,2$ mean that the order parameters involve fermions on electron pockets with either $xz (yz)$ or $xy$ orbital content, and indices $\Gamma$ and $M$
    mean that the SDW order parameter involves fermions from either $\Gamma-$centered or $M-$centered hole pockets.
    Using the values of the couplings along the FTs as inputs and solving these differential equations, we obtain
 $\Gamma_{SDW}^{(i)} \sim 1/(L_0-L)^{\beta_{SDW}^{(i)}}$, with $\beta_{SDW}^{(i)}=(1+\gamma_{3i}/a_i)/(1+\gamma_{3i}^2/a_i^2)$, where $i= (\Gamma,1;\Gamma,2;M,1;M,2)$ and $\gamma_{3i}\in\{\gamma_3,g_3,\gamma_{3n},g_{3n}\}$, $a_i\in\{a,a_n,c,c_n\}$. We verified that {\it all} $\beta_{SDW}^i$ are smaller than 1/2,  so that SDW order does not develop (if $L_0 < L_F$). The largest values are for $a=a_n=c=c_n=1$: $\beta_{SDW}^{\Gamma,1}=\beta_{SDW}^{\Gamma,2}=0.3$ and $\beta_{SDW}^{M,1}=\beta_{SDW}^{M,2}=0.43$.
  These are the same values  as in the toy model.

There are also four superconducting vertices: $\Gamma_{SC}^{e,xz/yz}, \Gamma_{SC}^{e,xy}, \Gamma_{SC}^{\Gamma}$, and $\Gamma_{SC}^{M}$.
The RG equations for these vertices can be cast into the matrix equation
\begin{align}\label{eq:SCfull}
\partial_L \bs{\Gamma}_{SC} =-2\begin{pmatrix}u_5+Hv_c & v_c+Hu_5 & u_3 & u_{3n} \\ v_c+Hv_5 & v_5+Hv_c & v_3 & v_{3n} \\ u_3+Hv_3 & v_3+Hu_3 & u_4 & u_c \\ \frac{u_{3n}+Hv_{3n}}{2} & \frac{v_{3n}+Hu_{3n}}{2} & \frac{u_c}{2}& \frac{u_{4n}}{2} \end{pmatrix}\bs{\Gamma}_{SC} ,
\end{align}
where we introduced $\bs{\Gamma}_{SC}=(\Gamma_{SC}^{e,xz/yz}, \Gamma_{SC}^{e,xy}, \Gamma_{SC}^{\Gamma}, \Gamma_{SC}^{M})^T$
 Along each FT the solution of Eq.~(\ref{eq:SCfull}) gives rise to  $s^{+-}$ gap structure on the contributing pockets.
  The exponents are
\begin{align}\label{eq:exponentsSCfull}
\beta_{SC}^{(4p_1)}&=\frac{-\gamma_4-\gamma_5+\sqrt{(\gamma_4-\gamma_5)^2+4\gamma_3^2}}{1+\gamma_{3}^2/a^2}\nonumber\\
\beta_{SC}^{(4p_2)}&=\frac{-\gamma_4-g_5+\sqrt{(\gamma_4-g_5)^2+4g_3^2}}{1+g_{3}^2/c^2}\nonumber\\
\beta_{SC}^{(3p_1)}&=\frac{-\gamma_{4n}-2\gamma_5+\sqrt{(\gamma_{4n}-2\gamma_5)^2+8\gamma_{3n}^2}}{1+\gamma_{3n}^2/a_n^2}\nonumber\\
\beta_{SC}^{(3p_2)}&=\frac{-\gamma_{4n}-2g_5+\sqrt{(\gamma_{4n}-2g_5)^2+8g_{3n}^2}}{1+g_{3n}^2/c_n^2}.
\end{align}
For $a=a_n=c=c_n$ we have $\beta_{SC}^{(4p_1)}=\beta_{SC}^{(4p_2)}=0.86$ and $\beta_{SC}^{(3p_1)}=\beta_{SC}^{(3p_2)}=0.72$, again as in the toy model.
We checked that $\beta_{SC}^{(i)}\geq1/2$  for all $a, a_n, c, c_n$, i.e., the superconducting susceptibility does diverge at $L=L_0$.

  To determine the SC gap structure on all pockets, we need to include the residual interactions. We did this numerically. We found that, like in the toy model,
   the largest eigenvalue in the SC channel corresponds to a "conventional'' $s^{+-}$  gap structure,
    although the magnitude of the gap on the "secondary'' pockets is small.
    Specifically, this means that  for 4p$_1$ the gap magnitude is relatively small on the $M-$centered hole pocket and the $xy-$part of the electron pockets, for 4p$_2$ it is small (very small) on the $M-$centered hole pocket and the $xz/yz-$parts of the electron pockets. In the 3p case, the gap almost vanishes on both $\Gamma-$centered hole pockets, and  the two 3p FTs differ in the gap magnitude on the $xz/yz$ and $xy$ portions of the electron pockets.

    For the second largest eigenvalue the gap structure for the FTs, where the dominant interactions are within the same orbitals (i.e.~4pFT$_1$ and 3pFT$_2$),
      is the orbital-antiphase $s^{+-}$ state, Ref.~\cite{comm_kot} (~sign$(\Gamma_{SC}^{e,xz/yz}, \Gamma_{SC}^{e,xy}, \Gamma_{SC}^{\Gamma}, \Gamma_{SC}^{M})=(+,-,-,+)$). For the FTs with dominant couplings between different orbitals (4pFT$_2$ and 3pFT$_1$) the  sign structure corresponds to ``orbital-antiphase $s^{++}$'' state (~sign$(\Gamma_{SC}^{e,xz/yz}, \Gamma_{SC}^{e,xy}, \Gamma_{SC}^{\Gamma}, \Gamma_{SC}^{M})=(+,-,+,-)$).


\begin{thebibliography}{10}
\bibitem{review}  Fernandes, R. M. Chubukov, A. V. and Schmalian, J.
 Nature Phys. 10, 97 (2014);  P. C. Canfield and S. L. Bud'ko, Annu. Rev. Condens. Matter Phys. 1, 27 (2010).
 \bibitem{rev_1}  Liang, S., Moreo, A. and Dagotto, E. Phys. Rev. Lett. 111, 047004 (2013).
 \bibitem{ch_rev} see, e.g., A.V. Chubukov, in "Iron-based Superconductivity", Springer Series in Materials Science, Vol. 211, pp. 255-329, (2015);
  Luca de' Medici, {\it ibid}
  pp. 409-441, (2015).
 \bibitem{we_last} A.V. Chubukov, M. Khodas, and R.M. Fernandes, arXiv:1602.05503.
 \bibitem{zlatko}   Cvetkovic, V. and Tesanovic, Z.,  Phys. Rev. B 80, 024512 (2009);
 \bibitem{orb_order} Yamase, H. and Zeyher, R., Phys. Rev. B 88, 180502(R) (2013); Lee, C. C., Yin, W. G. and Ku, W. Phys. Rev. Lett. 103, 267001 (2009);
Kruger, F. S., Kumar, J., Zaanen, J. and van den
Brink,  Phys. Rev. B 79, 054504 (2009);  Valenzuela, B., Bascones, E. and Calderon, M. J.Phys.
Rev. Lett. 105, 207202 (2010).  Lv, W. and Phillips, P., Phys. Rev. B 84,
174512 (2011); Lee, W-C. and Phillips, P. W. Phys. Rev.
B 86, 245113 (2012);  Applegate, R., Singh, R. R. P., Chen, C-C. and Devereaux,
T. P. Phys. Rev. B 85, 054411
(2012);  Stanev, V. and Littlewood, P. B., Phys. Rev. B 87, 161122(R)
(2013); Dumitrescu, P. T., Serbyn, M., Scalettar, R. T.,
and Vishwanath, A, arXiv:1512:08523 (2015);  Baek, S.-H., Efremov, D. V., Ok, J. M., Kim, J.
S., van den Brink, J. and Buchner, B. Nat Mater 14, 210 (2015); Gallais, I. and Paul, I. Comptes Rendus Physique
17, 113-139 (2016);   Wang, Z., and  Nevidomskyy,  A. H.,
Journal of Physics: Condensed Matter bf  27, 225602 (2015); Thorsmølle,  V. K., Khodas, M.,  Yin, Z.P.,  Zhang, C.,   Carr,  S.V.,  Dai, P.,  and  Blumberg, G.,  Phys. Rev. B 93, 054515 (2016)
\bibitem{FeSE}  Watson, M. D.,  Kim, T. K.,  Haghighirad, A. A., Davies, N. R.,   McCollam, A.,  Narayanan, A.,   Blake, S. F., Chen, Y. L.,  Ghannadzadeh,  S.,  Schofield, A. J.,  Hoesch, M., Meingast, C., Wolf, T. and  Coldea, A. I.,
 Phys. Rev. B 91, 155106 (2015); Y. Suzuki, T. Shimojima, T. Sonobe, A. Nakamura,
M. Sakano, H. Tsuji, J. Omachi, K. Yoshioka, M. Kuwata-Gonokami, T.
Watashige, R. Kobayashi, S. Kasahara, T. Shibauchi, Y. Matsuda, Y.
Yamakawa, H. Kontani, and K. Ishizaka, Phys. Rev. B 92, 205117
(2015);  Zhang, Y., \textit{et al}.
arXiv: 1503.01556;  Zhang, P. \textit{et al}. Phys. Rev. B 91, 214503
(2015); Kothapalli, K. \emph{et al},
arXiv:1603.04135 (2016); Fedorov, A, Yaresko, A, Kim, T. K., Kushnirenko,
E. Haubold, E, Wolf, T., Hoesch, M., Gruneis, A., Buchner, B., and
Borisenko S.  preprint.
\bibitem{parquet_RG} see e.g., A.T. Zheleznyak, V.M. Yakovenko, and I.E.
Dzyaloshinskii, Phys. Rev. B 55, 3200 (1997) and references therein.
\bibitem{fRG} Metzner, W., Castellani, C. and Di Castro, C.  Adv. Phys. 47, 317 (1998); Salmhofer M.,  Commun. Math. Phys. 194, 249 (1998).
\bibitem{fRG_thomalle} Platt, C., Honerkamp, C., and Hanke, W., New J. Phys. 11,
055058 (2009);
\bibitem{fRG_thomale2} Platt, C., Hanke, W. and Thomale, R.,
Advances in Physics 62, 453-562 (2013).
\bibitem{fRG_lee}  Yang, F., Wang, F., and Lee, D.-H.,
Phys. Rev. B 88, 100504 (2013).
\bibitem{rice} LeHur, K. and Rice, T. M.,  Ann. Phys. 324, 1452 (2009).
\bibitem{rahul} Nandkishore,L., Levitov,L., and Chubukov, A.V.,  Nature
Phys. 8, 158 (2012); Kiesel, M., Platt, C. Hanke, W., Abanin, D.A.,
and Thomale R., Phys. Rev. B 86, 020507 (2012).
\bibitem{cee} Chubukov, A. V., Efremov, D. V. and Eremin,
Phys. Rev. B 78, 134512 (2008)
\bibitem{pod} D. Podolsky, H-Y. Kee, Y. B. Kim, Europhysics Letters 88, 17004 (2009);   Maiti, S and Chubukov, A. V. Phys. Rev. B 82,214515 (2010).
\bibitem{kontani} Yamakawa, Y., Onari, S., and Kontani, arXiv:1509.01161
\bibitem{scalapino}  see. e.g., Kemper, A. F., Maier, T. A., Graser, S., Cheng,
H.-P., Hirschfeld, P. J. and Scalapino, D. J. New Journal of Physics 12, 073030 (2010) and references therein.
\bibitem{brouet} see, e.g.,
    V. Brouet, M. Fuglsang Jensen, Ping-Hui Lin, A. Taleb-Ibrahimi, P. Le Fèvre, F. Bertran, Chia-Hui Lin, Wei Ku, D. Colson, and A. Forget, Phys. Rev. B 86, 075123 (2012).
    \bibitem{sm_1} See Supplementary Material for detail.
    %AC
    The 40 interactions involve  pairs of fermions, each near either $\Gamma, X, Y$ or $M$ point.  There are 4 additional interactions involving fermions near each of these  points. These additional interactions %LC flow differently under RG and
    do not affect the behavior near the four stable fixed trajectioris that we found within the space of 40 couplings, as we explicitly verified. We neglect these additional interactions in our analysis.
    \bibitem{Cvetkovic2013}  Cvetkovic, V. and Vafek, O., Phys. Rev. B 88, 134510 (2013). See also Fernandes, R. M. and Vafek, O., Phys. Rev. B 90, 214514 (2014).
    \bibitem{maiti} see e.g. S. Maiti, A. Chubukov, Phys. Rev. B, 82, 214515 (2010) for a detailed explanation
    \bibitem{suscept_RG} A similar procedure has been used in the RG studies of other problems:
see, e.g.,   Metzner, W.,   Salmhofer, M.,  Honerkamp C.,  Meden, V., and  Schoenhammer K.,
 Rev. Mod. Phys. 84, 299 (2012)  and references therein;  Lemonik Y.,  Aleiner, I.L., and Fal'ko V.L.,  Physical Review B 85, 245451 (2012);
 Murray, J. M., and Vafek, O.,  Phys. Rev. B 89, 201110(R) (2014).
 \bibitem{eremin} Eremin, I. and Chubukov, A. V., Phys. Rev. B 81, 024511 (2010).
\bibitem{sdw_stripe} Chandra, P., Coleman, P., and Larkin, A. I., Phys. Rev. Lett. 64, 88, (1990);
Fang, C., Yao, H., Tsai, W-F., Hu, J. and Kivelson, S. A.  Phys. Rev. B 77, 224509 (2008); Xu, C., Muller, M., and Sachdev, S.,  Phys. Rev. B 78, 020501(R) (2008).
\bibitem{fkc} R.M. Fernandes, M. Khodas, and A.V. Chubukov, in preparation.
\bibitem{raf_last} M.N., Gastiasoro, I. Eremin, R.M. Fernandes, and B.M. Andersen, arXiv:1607.04711
\bibitem{rafael}  R. M. Fernandes, A.V. Chubukov, J. Knolle, I. Eremin, and J. Schmalian,  Phys. Rev. B 85, 024534 (2012).
%\bibitem{giamarchi} T. Giamarchi, {\it Quantum Physics in One Dimension} (Oxford, 2004).
\bibitem{comm_kot}  For the 3p case,   the subleading eigenfunction (smaller $\beta_{SC} >0$) describes the orbital-antiphase state with the gap sign on the $M$ pocket opposite to that on the other four pockets, see Yin, Z. P., Haule, K., and Kotliar, G., Nature Phys. 10, 845 (2014).
\bibitem{oo}
We call orbital order the symmetry breaking between $d_{xz}$ and $d_{yz}$ orbitals. Another $C_4$- symmetry breaking term is the difference in the
occupations of $d_{xy}$ orbitals at $X$ and $Y$ in the 1FeBZ~\cite{vafek_fern}. Such an order is present in our $3p_2$ model and $4p_2$ models.
 \bibitem{vafek_fern}R. M. Fernandes and O. Vafek, Phys. Rev. B 90, 214514 (2014).
\bibitem{we_future} For the application of the full 4p model to FeSe see R. Xing, L. Classen, M. Khodas, and A.V. Chubukov,  arXiv:1611.03912
\bibitem{comm_1} SC not preceded by SDW already at zero doping has been detected in a fRG analysis of a 4-pocket model and contrasted with the reported lack of such tendency in 5-pocket models~\protect\cite{ronny}.  We argue that the outcome of the pRG flow is qualitatively the same in both cases, only in the 5-pocket model the SC susceptibility overcomes the SDW susceptibility at smaller energies, i.e. after a longer RG flow.
\bibitem{ronny} R. Thomale, C. Platt, W. Hanke, B. A. Bernevig, Phys. Rev. Lett. 106, 187003 (2011).
\bibitem{vafek2016} O. Vafek and A.V. Chubukov, in preparation.
\end{thebibliography}
 \end{document}